# THz-Driven Coherent Phonon Fingerprints of Hidden Symmetry Breaking in 2D Layered Hybrid Perovskites

Joanna M. Urban[1]*, Michael S. Spencer[1], Maximilian Frenzel[1], Gaëlle Trippé- Allard[2], Marie Cherasse[1,3]‡, Charlotte Berrezueta Palacios[4], Olga Minakova[1], Eduardo Bedê Barros[5,6], Luca Perfetti[3], Stephanie Reich[4], Martin Wolf[1], Emmanuelle Deleporte[2], Sebastian F. Maehrlein[1,7,8]*

[1] *Fritz Haber Institute of the Max Planck Society, Department of Physical Chemistry, Faradayweg 4-6, 14195 Berlin, Germany*

[2] *Lumière, Matière et Interfaces (LuMIn) Laboratory, Université Paris-Saclay, ENS Paris-Saclay, CentraleSupélec, CNRS, 91190 Gif-sur-Yvette, France*

[3] *Laboratoire des Solides Irradiés, CEA/DRF/IRAMIS, École Polytechnique, CNRS, Institut Polytechnique de Paris, F-91128 Palaiseau, France*

[4] *Department of Physics, Freie Universität Berlin, Berlin 14195, Germany*

[5] *Department of Physics, Universidade Federal do Ceara, Fortaleza, Ceara, 60455-760 Brazil*

[6] *Institut für Festkörperphysik, Technische Universität Berlin, Hardenbergstraße 36, 10623 Berlin,Germany*

[7] *Helmholtz-Zentrum Dresden-Rossendorf, Institute of Radiation Physics, Dresden, Germany*

[8] *Technische Universität Dresden, Institute of Applied Physics, Dresden, Germany*

‡ Currently at Laboratoire d'Optique Appliquée, ENSTA Paris, CNRS, École Polytechnique, Institut Polytechnique de Paris, 91761 Palaiseau, France.

*email: maehrlein@fhi-berlin.mpg.de, urban@fhi-berlin.mpg.de

**Abstract**

**Metal-halide perovskites (MHPs) emerged as a family of novel semiconductors with outstanding optoelectronic properties for applications in photovoltaics and light emission. Recently, they also attract interest as promising candidates for spintronics. In materials lacking inversion symmetry, spin-orbit coupling (SOC) leads to the Rashba-Dresselhaus effect, offering a pathway for spin current control. Therefore, inversion symmetry breaking in MHPs, which are characterized by strong SOC, has crucial implications. Yet, in complex low-dimensional hybrid organic-inorganic perovskites (HOIPs), the presence of and structural**

**contributions to inversion symmetry breaking remain elusive. Here, employing intense THz fields, we coherently drive lattice dynamics carrying spectroscopic fingerprints of inversion symmetry breaking in Ruddlesden-Popper $(PEA)_2(MA)_{n-1}Pb_nI_{3n+1}$ perovskites, which are globally assigned to a centrosymmetric space group. We demonstrate coherent control by THz pulses over specific phonons, which we assign to either purely inorganic or highly anharmonic hybrid cage-ligand vibrations. By developing a general polarization analysis for THz-driven phonons, we pinpoint linear and nonlinear driving mechanisms. From this, we identify simultaneous IR- and Raman-activity of inorganic cage modes below 1.5 THz, indicating mode-selective inversion symmetry breaking. By exploring the driving pathways of these coherent phonons, we lay the groundwork for simultaneous ultrafast control of optoelectronic and spintronic properties in 2D HOIPs.**

## 1. Introduction

Metal halide perovskites (MHPs) have emerged over the past decade as novel semiconductors with exceptional photophysical properties, suitable for various optoelectronic applications, including photovoltaics and light emission.[1] Low-dimensional, hybrid organic-inorganic perovskites (HOIPs) offer enhanced stability compared to their 3D counterparts and vast structural tunability achieved via easy, solution-based synthesis.[2] Among them, layered 2D perovskites are promising candidates for light emitting diodes, lasers and photodetectors.[3] The remarkable properties of such low-dimensional HOIP compounds are largely influenced by electronic confinement of the charge carriers in the inorganic sublattice[4] as well as the significant coupling of the electronic and spin degrees of freedom to the vibrational excitations of the soft,[5] anharmonic,[6] highly polarizable[7] and dynamically disordered[8,9] lattice. Lowered dimensionality is reported to enhance the coherent lattice dynamics in HOIPs,[10] modify the phonon spectrum[11–13] and alter charge carrier-phonon coupling,[14] but the exact underlying mechanisms remain underexplored.

In addition, MHPs are recently intensely investigated as a new promising platform for spintronic applications.[15,16] In conjunction with a strong spin-orbit coupling (SOC) originating from the heavy metal atoms, lack of inversion symmetry in MHPs leads to the emergence of the Rashba-Dresselhaus effect:[17,18] a splitting of the electronic bands in momentum space, leading to an indirect bandgap and lifting of the spin degeneracy even in the absence of an external magnetic field. This makes inversion-symmetry-broken MHPs highly interesting for spin-charge conversion and spin-dependent charge transport applications. Beyond spintronics, inversion symmetry breaking also underlies phenomena such as ferroelectricity[19] or even-order optical nonlinearities.[20]

While the presence of inversion symmetry breaking in perovskite materials has far-reaching consequences, its exact nature in MHP materials often remains elusive and its presence and magnitude are debated.[21] Rashba-Dresselhaus splitting as a result of broken inversion symmetry has been reported in 2D layered Ruddlesden-Popper perovskites,[22–25] but the exact nature and conditions for the emergence of this effect remain debated. Significant research effort has recently been directed towards the chemical engineering of MHP materials with globally chiral structures, which would allow for controlling spin populations thanks to their chiroptical activity and exploring the chiral-induced spin selectivity (CISS) effect.[26] However, it has also been suggested that nontrivial, local inversion symmetry breaking can occur in HOIPs even in the absence of the breaking of global central symmetry,[27,28] leading to the emergence of the Rashba effect even in nominally centrosymmetric crystals.[29] The presence and origin of such intrinsic 'globally hidden' symmetry breaking in HOIPs remain controversial[30] and difficult to prove, especially in structurally complex, low-dimensional hybrid compounds, highlighting the need for more detailed studies and new experimental approaches. Especially, while most common characterization methods provide information about static material symmetries, the potential of time-resolved methods to study the dynamic character of the symmetry breaking at ultrafast timescales has not yet been fully explored. The analysis of optical signatures of phonon modes in crystalline materials offers sensitive insights into underlying lattice symmetries, which dictate the selection rules for vibrational excitations' interactions with light. Strong phonon-charge carrier coupling governs many of the MHP's remarkable optoelectronic properties, including surprising defect tolerance[31] and polaronic effects.[32] Vibrational modes could also mediate dynamic inversion symmetry breaking and lead to the emergence of a local, dynamical Rashba effect, as previously suggested for $MAPbI_3$.[33,34] Coherently driven phonons are therefore excellent handles for low-energy ultrafast material control, complementary to structural and chemical engineering, as recently demonstrated in 3D[7] and 2D[10] MHPs.

In this work, we therefore use intense THz fields to drive coherent vibrational modes in a family of layered 2D HOIP compounds with varying degrees of out-of-plane confinement (see **Figure 1a**). The increased structural ordering, enhanced by large aromatic PEA+ ligands, and the high quality of our phase-pure single crystal samples enable us to drive room temperature coherent lattice dynamics in Ruddlesden-Popper perovskite (RPP) compounds with $n$=1, 2, up to 3 inorganic metal-halide octahedral layers. This contrast to a previous study on a 2D perovskite system with short alkyl chain butylammonium ligands,[10] where clear coherent dynamics was only observed in the $n$=1 compound, highlighting the crucial role of specific ligand molecules and the interplay of the organic and inorganic sublattice, dictated by hydrogen bonding, steric effects, and electrostatic interactions.[35] Through a systematic study of THz-induced transient birefringence, we trace the evolution of lattice dynamics as a function of the number of the inorganic layers $n$, identifying a vibrational mode likely specific to the organic-inorganic sublattice interface. Our comprehensive analysis of mode symmetries and driving mechanisms identifies a globally hidden inversion symmetry breaking, evidenced by the simultaneous Raman and IR activity[36] of specific inorganic layer phonons. We demonstrate how tailoring excitation polarization conditions allows the selective targeting of specific vibrational modes, leveraging symmetry properties for ultrafast lattice-based material control.

## 2. Results

Here, we employ intense single-cycle THz pump pulses with peak fields of ~1 MV/cm (see Methods) and collinearly propagating 800 nm probe pulses, linearly polarized at 45° with respect to the THz, to stroboscopically sample the dynamics of the THz-induced birefringence [7,37] (see **Figure 1b**). The signal field is emitted in a wave mixing process by the nonlinear polarization induced via THz and visible/NIR field interactions.[7] The polarization component of the signal field perpendicular to the probe field is detected in a balanced detection scheme (see Figure 1b and **Supplementary Information, S17** and **S18**). The anisotropic signal generation in this nonlinear process can effectively be described as a rotation of the probe polarization by the angle $\phi$ due to transient changes of the refractive index (see **Figure 1b**). The probe photon energy is below the band gap of all the investigated samples, allowing us to investigate lattice dynamics in the unperturbed electronic ground state. With this method, we investigate single-crystal samples of lead iodide compounds from the family of layered Ruddlesden-Popper perovskites (RPPs) $(PEA)_2MA_{n-1}Pb_nI_{3n+1}$ (see **Figure 1a**, **Methods**, and **Supporting Information, S1**), where phenethylammonium cations ($PEA^+$) form the spacer bilayers and methylammonium cations ($MA^+$) are incorporated in the 12-fold coordination in between the octahedra in the inorganic sublattice for $n$=2, 3.

First, we measure the THz pump-induced transient birefringence for the $n$=1 sample at room temperature, varying the azimuthal crystal orientation in the plane perpendicular to the natural cleavage plane ($\psi$=0° azimuthal angle is defined as the long macroscopic crystal edge being parallel to the THz polarization, see inset **Figure 1c**). Both beams are incident normal to the sample surface, which corresponds to the 2D perovskite layer plane. Figure 1c shows the transient birefringence signal in $(PEA)_2PbI_4$ consisting of an instantaneous electronic response, which follows the square of the THz electric field, and a longer-lived oscillatory component. The THz-induced Kerr effect[37] (TKE) nature of the instantaneous signal is confirmed by the quadratic dependence of the instantaneous response peak at pump-probe delay $t = 0$ on the THz electric field (see **Figure 1d**). Furthermore, its amplitude shows a clear four-fold anisotropy (**Supporting Information, S7**), consistent with the four-fold pattern observed for visible-range nonlinear susceptibility in $(PEA)_2PbI_4$.[38] Based on this strong instantaneous nonlinear response, a nonlinear THz refractive index can be estimated as $n_2 \approx 1.4 \cdot 10^{-12} cm^2/W$ (lower limit, see **Supporting Information, S5**). This value is higher than for the related 3D compound $MAPbBr_3$ and several orders of magnitude higher than for many typical materials for nonlinear THz photonics.[7] This finding complements previous reports of very high optical nonlinearities of 2D perovskites in the visible frequency range.[39,40]

We assign the oscillatory part of the signal at longer times to coherent Raman-active phonon signatures which decay over the course of several picoseconds.[10] The Fourier transform (FT) of this oscillatory signal (calculated for $t > 1.5$ ps, **Figure 1e**) reveals multiple peaks in the 0.5–3 THz range, within the spectral bandwidth of the pump THz field and its nonlinear sum- and difference-frequency driving forces.[41] This strongly contrasts with the single low-frequency mode dominating the transient birefringence response of 3D MAPbBr$_3$,[7] which only becomes apparent at cryogenic temperatures due to strong phonon damping. Notably, a previous TKE study on 2D perovskites with butylammonium ligands revealed only a single vibrational mode at around 1.8 THz.[10] The larger number of Raman-active modes observed by us is likely due to the lower, triclinic crystal symmetry of (PEA)$_2$PbI$_4$ ($\overline{P1}$ space group[29,42]). The long-lived room temperature vibrational lattice coherence in our investigated 2D compound, significantly enhanced compared to the 3D MHPs, enables systematic studies of the phonon-modulated anisotropic transient birefringence response as a function of the sample azimuthal angle $\psi$. The coherence decay times of the most prominent modes, marked by the triangles in Figure 1e), are 3.5 ps, 1.5 ps, and 2 ps for the 0.9 THz, 1.7 THz, and 2.9 THz modes, respectively. These timescales agree with the estimated 2.4 ps decay time of the single 1.8 THz phonon mode observed in (BA)$_2$PbI$_4$ (Ref.[10]). Comparing the coherent phonon signals from several crystals of significantly different thicknesses allowed us to carefully exclude propagation effects as the origin of the long-lived oscillatory signals[7,43] (see **Supporting Information, S2**) and confirm their assignment as vibrational lattice modes.

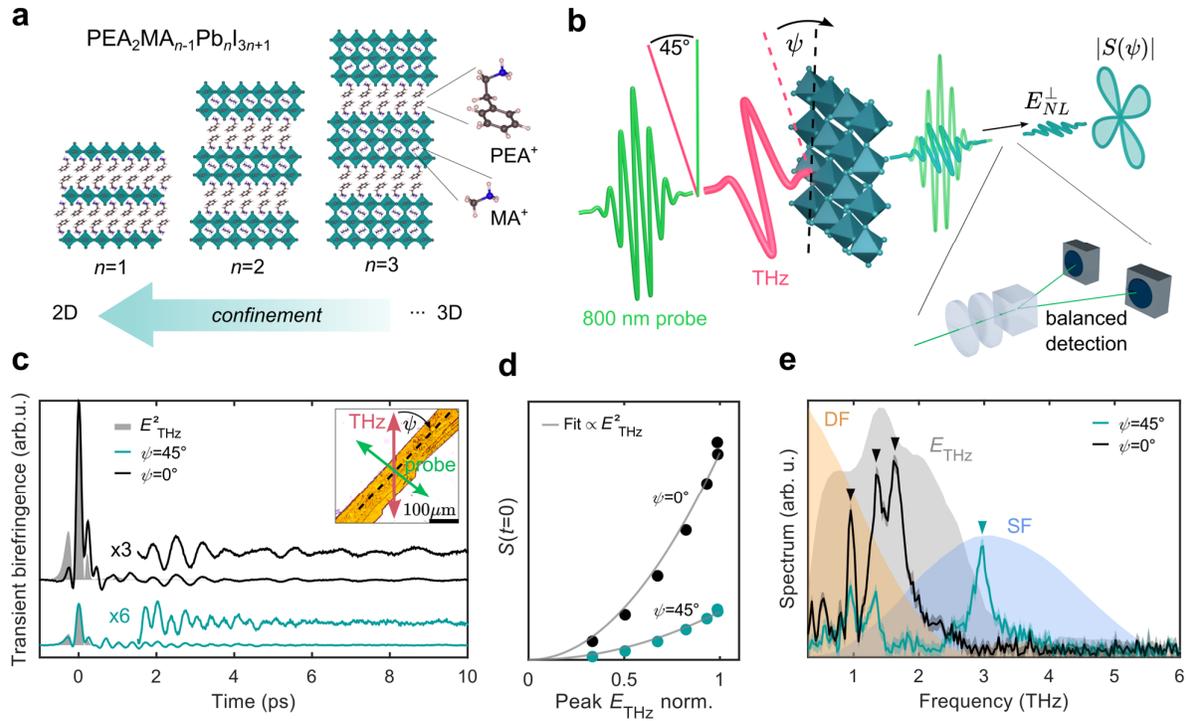

**Figure 1 a)** Schematic structure of Ruddlesden-Popper PEA$^+$-based perovskites. **b)** Schematic of the experimental configuration. The azimuthal angle-dependent signal $S(\psi)$ is proportional to $E_{NL}^\perp$, the signal field component polarized perpendicular to the initial probe polarization. **c)** RT transient birefringence traces for the $n$=1 sample for two different azimuthal crystal orientations. The instantaneous response approximately follows the square of the THz electric field measured by electro-optic sampling (shaded, grey). The enlarged traces show the oscillatory signal contribution. Inset: optical microscope image of the sample with the dashed line marking the long flake edge. **d)** Instantaneous response peak amplitude as a function of THz electric field for different sample orientations and corresponding quadratic fits. **e)** FTs of the oscillatory signal in panel c with dominant peaks marked, spectrum of the THz field (shaded grey) and the spectral profile of the difference- (orange) and sum- frequency (blue) nonlinear driving force.

Based on previous theoretical and Raman spectroscopy studies, we assign the low-frequency 0.9, 1.4 and 1.7 THz modes predominantly to vibrations of the PbI cage.[44,45] In 2D HOIPs, modes in this frequency range have been previously identified as tilts and rotations of the octahedral cages[44–46] (~0.8 THz), predominantly in-plane[44] metal-halide bond bending and stretching (1.3-1.7 THz)[44,46–48] as well as rattling and torsional modes of both the large ligand molecules and the smaller organic cations inside the inorganic cage (0.8 and 1.7 THz).[48,49] The 1.7 THz mode is close in frequency to the single dominating 1.8 THz Raman-active phonon in (BA)$_2$PbBr$_4$ assigned to the inorganic cage.[10] The relative shift in frequency between the two compounds is attributed to the larger mass of the iodide compared to bromide ions. Driving these low-frequency metal halide cage modes is key to ultrafast control of RPP optoelectronic properties, as they are known to strongly modulate the electronic structure of (PEA)$_2$PbI$_4$ and related compounds.[50,51] The mode around 2.9 THz in 2D HOIPs has previously been assigned to either out-of-plane metal-halide stretching[46] or a hybrid vibration involving the organic ligand cation.[47] In contrast, we did not observe this mode in 3D MAPbBr$_3$,[7] and only a very short-lived 2.9 THz vibration was reported in MAPbI$_3$ in a recent THz-driven broadband optical spectroscopy study,[52] therefore the prominence of this phonon appears to be characteristic to our 2D system. Given the evidence of enhanced lattice coherence times in the $n$=1 PEA$^+$-based RPPs compared to 3D HOIP compounds and the appearance of multiple previously unreported vibrational modes, we next investigate the dependence of the lattice dynamics on the degree of confinement by comparison with $n$ = 2, 3 samples.

Transient birefringence measurements at the room temperature are compared in **Figure 2a** for $n$ = 1, 2 and 3 for crystal edge orientation at 45° to the THz polarization direction (see also **Supporting Information, Fig. S1**). Here, we witness clear coherent phonon signatures for $n$ = 2, 3, in contrast to previously investigated butylammonium/methylammonium (BA/MA)-based RPP compounds, which only displayed coherent dynamics for $n$=1.[10] We attribute this difference to a higher structural ordering imposed by the bulky, stiff PEA$^+$ cations which can undergo $\pi - \pi$ stacking.[53] **Figure 2b** shows the corresponding FT amplitude spectra of the oscillatory signals (for $t > 1.5$ ps), for two crystal orientations $\psi = 0°$ and $\psi = 45°$. Coherent phonon dynamics in pure PEA$^+$ ligand-halide salt crystals (see **Supporting Information, S15**) significantly differ from those in the perovskite crystals, confirming the observed lattice dynamics in RPPs as an emergent property of the entire organic-inorganic structure.[48]

As shown in Figure 2b, we observe a clear amplitude decrease of the 2.9 THz mode relative to the low-frequency cage modes with an increasing number of inorganic layers from $n$=1 to $n$=3. We therefore tentatively assign the 2.9 THz to altered vibrational dynamics at the interface between the organic spacer ligands and the inorganic cage layers. The decrease in the mode's relative amplitude with increasing number of layers $n$ would in this case arise from a reduced dominance of the organic-inorganic interface. The mode could be related to rigid-body dynamics of the PEA$^+$ cation[47,48] or to inorganic cage vibrations activated in the altered local environment of the outermost inorganic octahedral layer interfacing the ligands.[48] Recently reported ab initio calculations predict higher mobility and mean displacements of halogen ions at the interface, leading to significantly larger vibrational density of states above 2.5 THz compared to the the intralayer ions, supporting this interpretation.[54] Alternatively, it could be a purely inorganic cage mode, more strongly damped in the higher-$n$ and 3D compounds due to the presence of MA$^+$ cations.[10,55]

To gain further insight into the vibrational modes' nature, we characterize the lattice dynamics for the $n$ = 1 sample as a function of temperature. **Figure 2c** and **d** show the FTs of the transient birefringence signals in the temperature range from 80 K to 295 K. The lower-frequency modes (**Figure 2c,** $\psi = 0°$) and the 2.9 THz mode (**Figure 2d**, $\psi = 45°$) slightly soften and broaden with increasing temperature.

However, no significant spectral changes are observed. This confirms the absence of phase transitions in this temperature range and is consistent with previous Raman spectroscopy studies.[42] The softening of vibrational modes with temperature is a result of lattice anharmonicity[56] which underlies both lattice expansion and phonon-phonon scattering (see **Supporting Information, S10**). We plot the relative shift of the four characteristic modes in **Figure 2e** along with predictions of a simple model considering only a symmetric anharmonic decay into two lower-frequency phonons[57] (inset of **Figure 2f,** see Methods**)**. Strikingly, the 2.9 THz mode shifts significantly more with temperature than all the low frequency modes, suggesting the presence of more efficient anharmonic decay channels. The temperature dependence of the extracted phonon lifetimes (see Figure 2f) is compared to a model including defect scattering[58] and temperature-dependent phonon-phonon scattering[56] (see **Supporting Information, S10**). This simple model qualitatively captures the observed evolution and highlights the much higher anharmonicity of the 2.9 THz mode compared to the lower frequency modes. At low temperatures, the 2.9 THz mode has a longer coherence time than the inorganic cage modes ($\tau_{1.4\text{THz}} \approx 1.7$ ps, $\tau_{1.7\text{THz}} \approx 2$ ps, $\tau_{2.9\text{THz}} \approx 3$ ps at 80 K). At room temperature, the situation is reversed and the 2.9 THz mode is more strongly damped ($\tau_{2.9\text{THz}} \approx 1$ ps) while the inorganic mode lifetimes are only very slightly affected by temperature (as can also be seen comparing the relative linewidths of the 0.9 THz and 2.9 THz modes at RT in Figure 1e). In line with our tentative assignment of this mode to a hybrid ligand-inorganic vibration, we propose that the strong decrease of its lifetime with temperature is due to scattering with thermally activated vibrations at the organic-inorganic interface. The lack of direct bonds between the organic and inorganic layers in RPPs typically results in significantly higher anharmonicity compared to lead organic chalcogenides with direct inorganic-ligand connectivity.[59] The stability of hydrogen bonds between the ligand molecules and halide ions in HOIPs strongly decreases with temperature,[60] potentially increasing dynamic disorder and reducing the lattice coherence. Weakening of the non-covalent interactions mediating the organic-inorganic sublattice coupling may explain the softening of a hybrid organic-inorganic mode at higher temperatures. The different scattering rates for the specific vibrations may also reflect distinct selection rules for the scattering processes imposed by local symmetries of the respective lattice subsystems.[59] The observed nontrivial evolution of phonon amplitudes and lifetimes is a signature of strong anharmonicity and nonlinear phonon-phonon coupling in hybrid 2D layered perovskites.

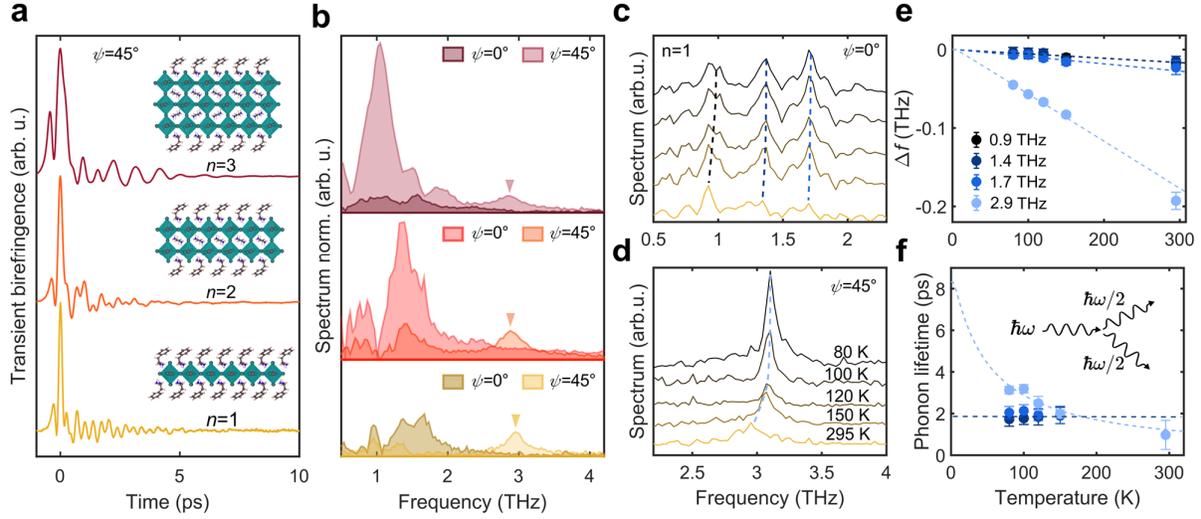

**Figure 2 a)** RT transient birefringence traces for RPPs $n$=1, 2, 3 normalized to the instantaneous peak amplitude. **b)** FTs of the oscillatory signal for the RPPs at $\psi = 45°$ and $\psi = 0°$ crystal edge orientation with the 2.9 THz mode marked. **c)** Temperature dependence of the spectra for $n$=1 measured at $\psi = 0°$ showing the low frequency cage modes. The dashed lines marks the evolution of the characteristic peaks. **d)** Temperature dependence of the spectra for $n$=1 measured at $\psi = 45°$, showing the evolution of the 2.9 THz mode. **e)** Relative frequency shift of the phonon modes with temperature and predictions of a 3-phonon scattering model for the different modes (dashed lines). **f)** Lifetimes of the modes as a function of temperature, estimates by a 3-phonon scattering model for the 2.9 THz mode (light blue dashed line) and approximately temperature-independent lowest-frequency mode lifetime (dark blue dashed line). Inset: schematic of symmetric 3-phonon scattering.

To further investigate the intriguing anisotropy of the transient birefringence signal and the driving mechanisms of the coherent lattice modes, we perform measurements as a function of crystal orientation relative to the pump and probe polarization directions for the $n = 1$ sample. In this way, we can systematically analyze the symmetries of the phonon modes. In **Figure 3a** we map the measured spectral amplitude as a function of rotation angle $\psi$ in the inorganic layer plane. The four dominant modes around 0.9, 1.4, 1.7 and 2.9 THz exhibit distinct angular signatures. **Figure 3b** shows the clear quadratic dependence of the 1.7 and 2.9 THz mode amplitude on the THz electric field. This scaling suggests a nonlinear driving mechanism, which could be either Raman-type photonic excitation,[41,61] where two THz photons within the bandwidth of the broadband THz pulse excite a Raman-active vibration $\Omega_R$ in a sum-frequency[41] or difference-frequency process (as shown in **Figure 3c**), or ionic Raman-type excitation, via the anharmonic coupling of two directly driven IR-active phonons.[61] The direct (i.e. linear) excitation pathway remains forbidden for a Raman-active mode in the absence of inversion symmetry breaking, due to the rule of mutual exclusion, which prohibits simultaneous IR- and Raman-activity.[62]

In the following, we show how the polar patterns observed for the dominant modes (shown in **Figure 3d** and **e**) can be explained in the framework of the nonlinear photonic driving for the 1.7 and 2.9 THz phonons, and how they allow us to unveil the specific mode symmetries. To do this, we separately consider the Raman-type coherent phonon driving and probing processes (see **Figure 3c**). The measured signal amplitude $S$ is proportional to the product of the driving force $F_{dr}^R$ and probing sensitivity $S_{pr}$, i.e. $S \propto S_{pr} F_{dr}^R$. In the two-photon excitation scheme, the coherent phonon at frequency $\Omega_R = \omega_{THz,1} \pm \omega_{THz,2}$ is driven by the interaction of the system with two THz fields $\boldsymbol{E}^{THz,1}$ and $\boldsymbol{E}^{THz,2}$ at frequencies $\omega_{THz,1}$ and $\omega_{THz,2}$, which perturb the electronic cloud and create a driving force $F_{dr}^R$ along the vibrational mode coordinate. The driving force is proportional to (see details in **Supporting Information, S17**):

$$F_{dr}^R \propto \boldsymbol{E}^{THz,2} \cdot \mathbf{R}^{\omega_{THz,1}, \Omega_R} \boldsymbol{E}^{THz,1}, \tag{1}$$

where $\mathbf{R}^{\omega_{THz,1}, \Omega_R}$ is the Raman susceptibility tensor. The values of the tensor components are dictated by the $\Omega_R$ mode symmetry and generally depend on the driving field frequencies $\omega_{THz,1}$ and $\omega_{THz,2}$. The probing is also described as a Raman-type process (as shown in Figure 3c), in which the probe field $\boldsymbol{E}^{pr}$ is scattered by the coherent lattice vibrations, which gives rise to Stokes and anti-Stokes sidebands. The scattered field components polarized perpendicular to the probe field's polarization are detected and give rise to the signal (see **Supporting Information S17** for full description). For simplicity, only the anti-Stokes probing pathway is shown schematically in Figure 3c. The probing sensitivity $S_{pr}$ is proportional to $S_{pr} \propto \hat{\boldsymbol{e}}_{pr}^{\perp} \cdot \mathbf{R}^{\omega_{pr}, \Omega_R} \boldsymbol{E}^{pr}$, where $\mathbf{R}^{\omega_{pr}, \Omega_R}$ is the Raman tensor of the mode, defined for the incident probe field frequency $\omega_{pr}$ (see Supporting Information), and $\hat{\boldsymbol{e}}_{pr}^{\perp}$ a unitary vector perpendicular to the probe polarization.

The Raman-active modes in crystals belonging to the triclinic $\overline{P1}$ space group have $A_g$ symmetry. In principle, all Raman tensor elements for such modes can be nonzero. We note, however, that only the anisotropic contributions to the Raman tensor elements give rise to the scattered field components which are detected in our balanced detection scheme. We find that the polar dependence of the TKE signal for the 1.7 THz and 2.9 THz modes can be very well reproduced based on tensors obtained by fitting the polar dependence of their intensity in polarization-resolved spontaneous Raman scattering experiments (see **Supporting Information S14**) by:

$$\mathbf{R}^{\alpha} = \begin{bmatrix} a & \frac{4}{3} a e^{i\pi/2} \\ \frac{4}{3} a e^{i\pi/2} & -a \end{bmatrix}, \qquad \mathbf{R}^{\beta} = \begin{bmatrix} b e^{i\pi/2} & -\frac{8}{9} b \\ -\frac{8}{9} b & -b e^{i\pi/2} \end{bmatrix}. \tag{2}$$

To describe the probing at 800 nm we assume tensors $\mathbf{R}^{\omega_{\text{pr}},\Omega_R}$ identical to the ones obtained from fitting the spontaneous Raman scattering data, and use the real part of these tensors as $\mathbf{R}^{\omega_{\text{THz,1}},\Omega_R}$ to construct the driving force $F_{\text{dr}}^R \propto \mathbf{E}^{\text{THz,2}} \cdot \mathbf{R}^{\omega_{\text{THz,1}},\Omega_R} \mathbf{E}^{\text{THz,1}}$. We calculate the predicted azimuthal angle dependence for the 1.7 THz and 2.9 THz mode amplitudes, using $\mathbf{R}^\alpha$ and $\mathbf{R}^\beta$ and their real parts respectively, as shown with black lines in **Figure 3d.** The deviations from the calculated pattern are predominantly artifacts due to inhomogeneities of the rotating sample (see also **Supporting Information, S7**). We summarize this mechanistic concept for the 2.9 THz mode in **Figure 3e**, which shows the calculated azimuthal angle dependence of the driving force $|F_{\text{dr}}^R|$ and the probing sensitivity $|S_{\text{pr}}|$ amplitudes based on $\mathbf{R}^\beta$. Their product matches very well the experimentally extracted 2.9 THz mode amplitude variation. Performing the same calculation using a tensor of the form $\mathbf{R}^\alpha$ allows us to reproduce the 1.7 THz mode's behavior.

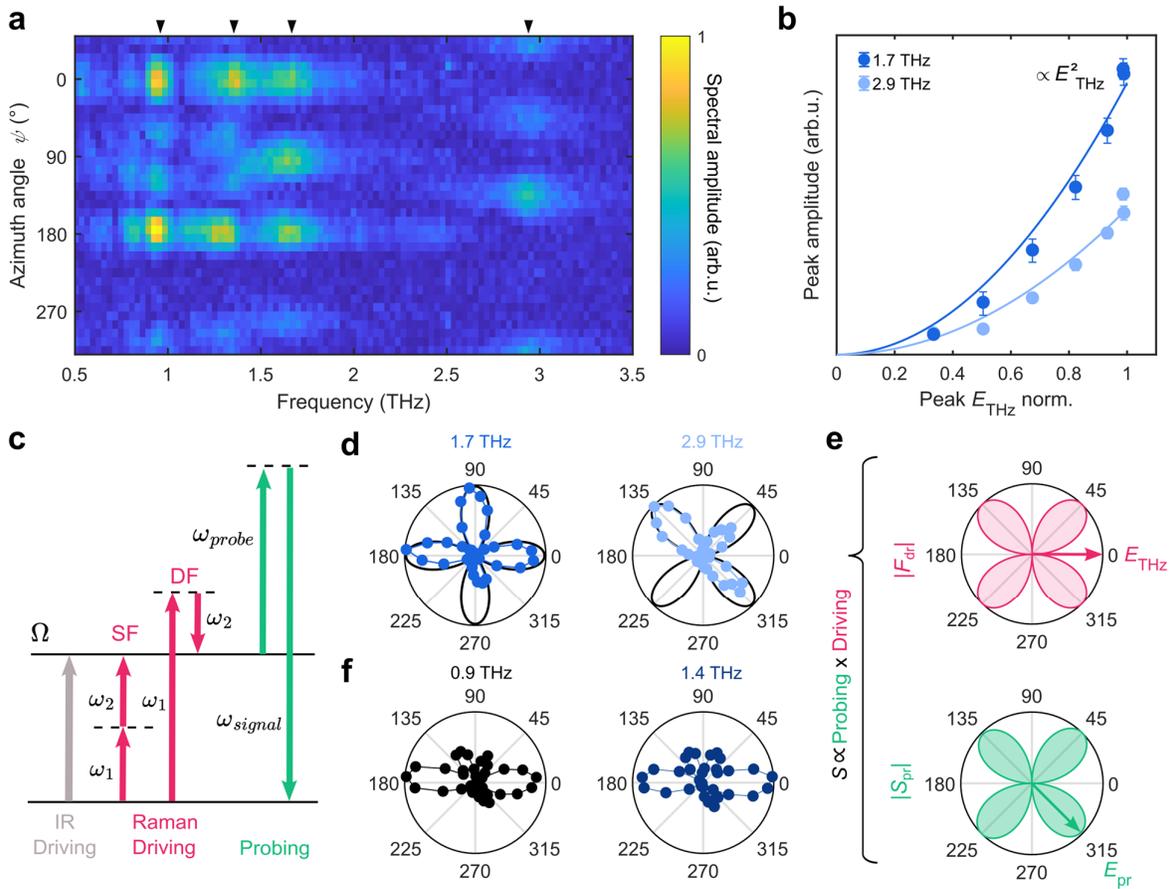

**Figure 3 a)** Azimuthal angle dependence of the spectral amplitude of the transient birefringence FT. $\psi = 0°$ corresponds to the crystal long axis oriented parallel to the THz electric field. **b)** Amplitude of the 1.7 and 2.9 THz modes as a function of the peak THz electric field and quadratic fits **c)** Schematic of nonlinear difference- and sum-frequency Raman excitation driving by THz fields with frequencies $\omega_1$ and $\omega_2$ and Raman-type probing mechanism **d)** Amplitude of the 1.7 THz and 2.9 THz FT peaks as a function of the sample's azimuthal angle (points) and theoretically calculated TKE signal amplitudes (black solid line). **e)** Amplitude of Raman-type driving force $|F_{\text{dr}}|$ and detection efficiency $|S_{\text{pr}}|$ as a function of sample azimuthal rotation angle. Arrows indicate the polarization directions of the THz and incident probe electric fields. f) Amplitude of the 0.9 and 1.4 THz modes as a function of sample azimuthal angle.

In striking contrast, the 0.9 THz and 1.4 THz modes show a six-lobed polar amplitude pattern (see **Figure 3f**), which cannot be modelled well assuming a Raman-type driving and detection mechanism (see **Supporting Information, S12**). In the following, we show that this experimental result can be explained by assuming simultaneous IR and Raman activity of these modes, which is a clear sign of broken inversion symmetry.[62] To decode the peculiar signal symmetry, we make use of the phase information provided by the coherent nature of our time-domain method and analyze the full complex FT. **Figure 4a** shows the imaginary part of the FT of the signal as a function of the azimuthal angle $\psi$ to illustrate the phase changes of the oscillatory phonon signal with sample rotation. A sample rotation by 180° is equivalent to inverting the THz field polarity ($\boldsymbol{E}^{\mathrm{THz}} \to -\boldsymbol{E}^{\mathrm{THz}}$) relative to the crystal. Comparing the inverse FTs of the frequency filtered 0.9 and 1.7 THz modes (see box masks in **Figure 4a**) in **Figure 4b** we observe a clear $\pi$ phase shift of the oscillatory signal for the 0.9 THz mode and an unchanged phase of the 1.7 THz mode upon 180° rotation. The $\pi$ phase shift corresponds to a change of sign of both the real and imaginary part of the complex FT, as seen in the phase map of **Figure 4a** and demonstrated in **Figure 4c**.

To rationalize this observation, we assume a linear driving via the coupling of the THz field to the electric dipole moments of IR-active 0.9 and 1.4 THz modes. In contrast to the Raman-type driving mechanism discussed before, the sign of the linear driving force depends on the THz electric field's polarity and $F_{\mathrm{dr}}^{\mathrm{IR}}(\psi = 0°) = -F_{\mathrm{dr}}^{\mathrm{IR}}(\psi = 180°)$. The assumption of a linear driving mechanism is confirmed by the linear scaling of the low-frequency modes' amplitude with the THz peak field (**Figure 4d**). The observation is also consistent with the broad excitation spectrum covering the 0.9 and 1.4 THz frequencies and with complementary THz transmission experiments (see **Supporting Information, S16**).

Finally, we explain the observed polar pattern by considering the decomposition of the signal amplitude into excitation and detection terms in an IR-drive-Raman-probe type experiment.[36] We find the dipole moments of the IR-active modes to be oriented along the long axis of the flake ($\psi = 0°$). The driving force is proportional to the projection of the THz electric field along the mode dipole moment, which can be expressed as:

$$F_{\mathrm{dr}}^{\mathrm{IR}} \propto \boldsymbol{\mathcal{Z}}^* \cdot \boldsymbol{E}^{\mathrm{THz}}, \qquad (3)$$

where $\boldsymbol{\mathcal{Z}}^*$ is the mode effective charge. Thus, the driving force amplitude follows a dipolar pattern and is maximal for THz parallel to the flake long axis, i.e. $\psi = 0°$ or 180°. As our detection scheme is only sensitive to Raman-active modes, the 0.9 and 1.4 THz linearly driven modes must simultaneously be both IR and Raman active.[36] We calculate the probing sensitivity for those modes as described previously, using a Raman tensor of the form $\mathbf{R}^\alpha$. The product of the probing sensitivity and driving force reproduces well the observed azimuthal angle dependence of the signal in **Figure 4e** and **4f**. Not only the amplitude (**Figure 4e**), but also the sign of the theoretically calculated signal (**Figure 4f**) agrees well with the experimental data, reflecting the $\pi$ phase jump between neighboring lobes of the polar pattern.

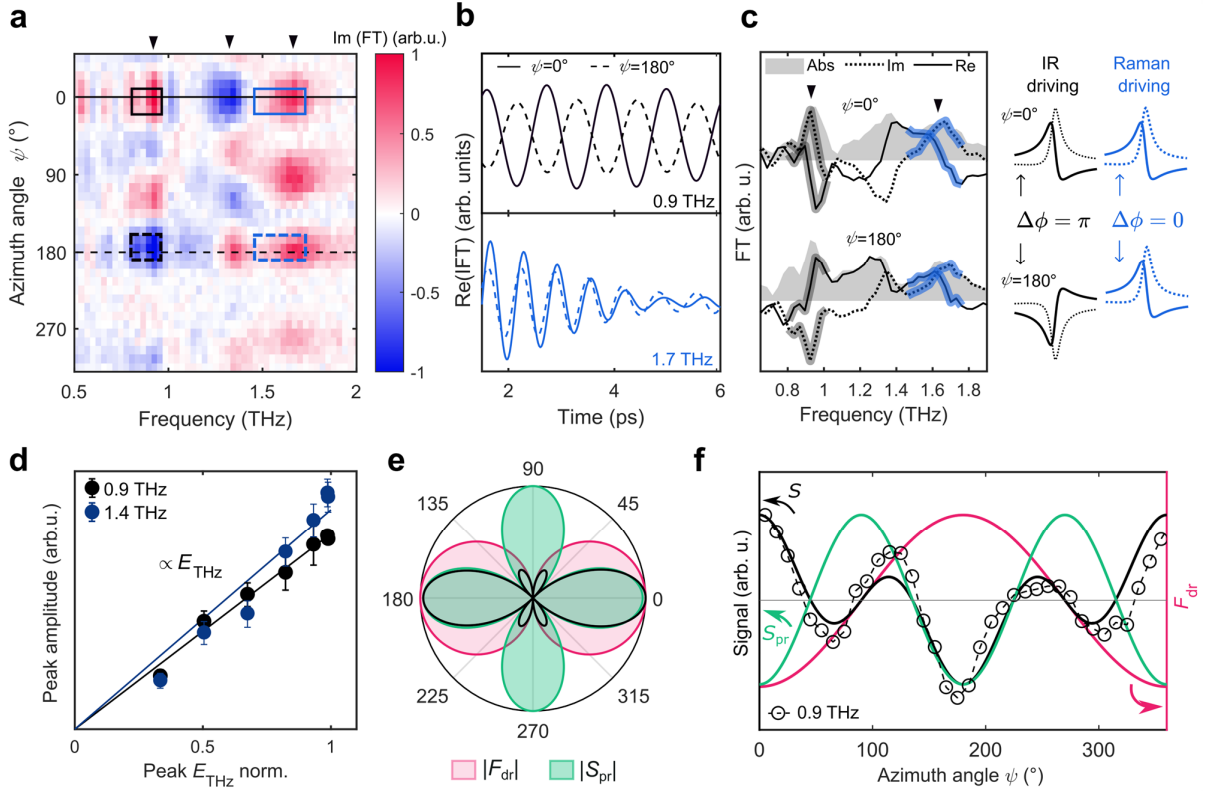

**Figure 4 a)** Normalized imaginary part of the FT of the transient birefringence signal as a function of the sample azimuthal angle. The boxes marked with solid (dashed) lines indicate regions in which the signal was averaged around $\psi = 0°$ ($\psi = 180°$) for calculating the inverse FT. **b)** Inverse FTs of the masked regions in panel a), showing the time domain signal corresponding to the 0.9 THz and 1.7 THz phonon modes at $\psi = 0°$ (solid) and $\psi = 180°$ (dashed). **c)** The amplitude (shaded gray area) and real (solid) and imaginary (dotted) part of the FT around $\psi = 0°$ and $\psi = 180°$ with highlighted spectral regions used to obtain traces shown in (b). Sketches on the right show the behavior of the complex FT upon 180° sample rotation for an IR- and Raman-driven resonance. **d)** Amplitude of the 0.9 THz and 1.4 THz modes as a function of the THz field and linear fits. **e)** Amplitude of the driving force (shaded magenta), probing sensitivity (green) and detected transient birefringence signal (black) as a function of the azimuthal angle **f)** Azimuthal angle dependence of the 0.9 THz mode amplitude, multiplied by the sign of the FT imaginary part to illustrate the changes of the phase (circles), the theoretically calculated transient birefringence signal $S$ (black), driving force $F_{\mathrm{dr}}^{\mathrm{IR}}$ (magenta) and probing sensitivity $S_{\mathrm{pr}}$ (green).

The dominance of linear excitation via linear THz field coupling to the mode's dipole moment (mode effective charge) for modes, which are both IR- and Raman-active, can be explained by the expected lower efficiency of the nonlinear driving pathway. Crucially, the simultaneous IR and Raman activity of the 0.9 THz and 1.4 THz modes contradicts the mutual exclusion principle imposed by a globally centrosymmetric crystal structure. We observe the same phase behavior for the low-frequency cage modes in $n$=2 and 3 samples (see **Supporting Information, S13**), indicating that globally hidden local or dynamic symmetry breaking is experienced by phonon modes of the inorganic sublattice in the entire RPP family.

**3. Discussion**

Using intense THz fields, we drive coherent lattice dynamics in compounds of the $(PEA)_2(MA)_{n-1}Pb_nI_{3n+1}$ 2D perovskite family. In contrast to BA$^+$-based RPP compounds, which previously showed room temperature coherent lattice responses only for n=1 inorganic layers,[10] we observe the excitation of coherent phonons also for the less confined $n$=2,3 structures, extending the potential for phonon-driven material control under application-relevant conditions. This difference in lattice coherence when employing PEA$^+$ and BA$^+$ ligands highlights the impact of the organic cation on the global lattice behavior. The highly hydrophobic, bulky PEA$^+$ cations improve material stability[63] and lead to reduced structural disorder, which is likely the key to the observed extension of phonon coherence times.[58] We furthermore identify a strongly anharmonic 2.9 THz mode that is absent in related 3D compounds,[7] which we assign to a hybrid organic-inorganic interface mode and which potentially mediates the organic-inorganic sublattice coupling and could be used as a handle to target interlayer charge transfer[64] in 2D RPPs. Our observations confirm the strong nonlinear response of layered 2D perovskites in the THz range,[10] complementing previous reports of exceptionally high nonlinearities in the optical range,[39] suggesting application potential for ultrafast THz photonic devices.

We observe multiple anisotropic Raman-active modes in (PEA)$_2$PbI$_4$ in the 0.5-3 THz range and with our systematic polarization- and phase-resolved analysis, we identify their linear and nonlinear excitation conditions, laying the groundwork for selective THz lattice control. Strikingly, the two lowest-frequency modes at 0.9 and 1.4 THz show simultaneous IR- and Raman-activity. In the closely related (BA)$_2$PbI$_4$ compound, a THz transmission study found 0.9 and 1.2 THz IR-active transitions with clear linear in-plane anisotropy polarized along the same crystal direction, assigned as octahedral rocking modes.[65] Given the structural similarity to (PEA)$_2$PbI$_4$, the two lowest frequency modes observed by us are likely related to analogous inorganic cage dynamics. A local inversion symmetry breaking is a plausible explanation for the intriguing lifting of the mutual exclusion rule. With our broadband single-cycle THz pulse, covering 0.5-2.5 THz, we expect to efficiently linearly drive all of the IR-active modes in this range. The simultaneous Raman and IR activity could alternatively be explained by a modification of the optical selection rules by strong spatial electric field gradients on the length scales of atomic bonds, as in the case of surface-enhanced Raman scattering experiments.[66] However, these should not occur under our experimental conditions. Interestingly, the linearly- and nonlinearly-driven modes show similar amplitudes in the transient birefringence signal. This is surprising, since signal amplitudes observed in a second-order nonlinear process (linear driving and Raman probing) are expected to be much higher than for a third-order (Raman driving and Raman probing) process. Previous reports of weaker second harmonic generation signals compared to third harmonic signals in (PEA)$_2$PbI$_4$[20] are in agreement with this observation.

The topic of inversion symmetry breaking in HOIPs has been widely studied using second harmonic generation (SHG) spectroscopy and microscopy.[20,67] In PEA$^+$-based RPPs, SHG has been previously observed and attributed to symmetry breaking induced by the ligand molecules.[20] However, reports on the presence and absence of SHG in samples of nominally identical composition are often inconsistent. The discrepancies have been assigned to edge states, grain boundaries and cancellation from random grain orientation.[20] These factors should be negligible in our study on highly ordered single crystals. However, we did not observe any SHG in macroscopic experiments using 800 nm pump/400 nm detection on the samples studied here. SHG-based studies can further be complicated by multiphoton photoluminescence,[20] highlighting the advantages of our coherent phonon fingerprint approach to detecting inversion symmetry breaking. While SHG measurements primarily are sensitive to the electronic nonlinear susceptibility, our method directly interrogates these specific ionic motions, which witness an inversion symmetry-broken lattice potential. Our approach offers an alternative to conventional THz transmission spectroscopy for studying IR-active modes in non-centrosymmetric materials.[36]

The inversion symmetry breaking observed here is unexpected, as most structural studies of (PEA)$_2$PbI$_4$ determine a globally centrosymmetric, triclinic $\overline{P1}$ space group.[29,42] However, crystallographic studies often provide incomplete or averaged information on the organic cation orientation, potentially leading to conflicting assignments of the same compound to either centrosymmetric or non-centrosymmetric space groups.[20] Two recent works based on combined X-ray diffraction and DFT analysis proposed a multiconfigurational, polymorphic ground-state structure in (PEA)$_2$PbI$_4$, leading to intrinsic inversion symmetry breaking and the presence of the Rashba effect, despite the structure being conventionally assigned as centrosymmetric.[29,68] The tilt of the inorganic cages around the $c^*$ crystallographic axis and the rotation of the ligand molecules were proposed as degrees of freedom responsible for the emergence of the different phases. Inversion symmetry breaking could emerge at the interfaces between the distinct sub-phases, each belonging to a centrosymmetric $\overline{P1}$ space group,[29] or occur in entire crystalline domains due to misalignment of the PEA$^+$ cations in adjacent layers.[68]

In our work, the lowest-energy Pb-I modes serve as dynamic fingerprints of inversion symmetry breaking, independent of the number of the octahedral layers ($n$=1-3). This contrasts with mechanisms related to the out-of-plane symmetry breaking dependent on an odd/even number of inorganic layers recently proposed for RPPs.[22] Thus, we tentatively attribute our observation of hidden symmetry breaking to local rotational distortions of the inorganic cages, present independent of the out-of-plane layer stacking and ordering of the layers. In-plane phonons related to inorganic cage tilting would be natural indicators of such nontrivial, intrinsic symmetry breaking.

In complex structures like layered 2D HOIPs, it is essential to consider symmetries separately at the octahedral cage, inorganic sub-unit, and on a global level.[27] Several recent works proposed the presence of local, as opposed to global inversion symmetry breaking in MHPs.[27,28] The presence of local inversion-symmetry breaking, despite a globally centrosymmetric space group, has been previously linked to structural disorder on the micrometer scale in polycrystalline 3D HOIPs.[67] In our highly ordered single crystals, the local symmetry breaking may alternatively also be related to quasi-one-dimensional twin domain boundaries within the RPP layers, arising due to collective alignment of the organic molecules and lead-halide cage distortions, as recently reported for $BA^+$-based RPPs.[69,70] Previously, transient dynamic inversion symmetry breaking mechanisms, related to instantaneous distortions of the lead-halide local environment,[33,71,72] as opposed to static distortions, were proposed in HOIPs. At this point, however, we cannot distinguish between the static or dynamic origin of the observed effect. Given that phonons at frequencies as low as 0.9 THz bear the signatures of the broken inversion symmetry, we expect the symmetry breaking to persist for at least the time of an oscillation cycle, that is > 1 ps. Regardless of its static or dynamic character, the inversion symmetry breaking evidenced in our work is highly consequential for the optoelectronic and spintronic properties of the $(PEA)_2MA_{n-1}Pb_nI_{3n+1}$ perovskite family, potentially leading to the emergence of a Rashba effect, chiro-optical activity and ferroelectricity.

Exploring the exact microscopic nature and potentially spatially localized character of the hidden inversion symmetry breaking opens an exciting avenue for future research using spatially-resolved methods such as sum-frequency microscopy,[73] Kerr effect microscopy,[74] THz s-SNOM[75] or piezoresponse force microscopy.[76] Composition tuning of the perovskite materials, for example through ligand engineering,[77] could be employed to control the inversion symmetry breaking, as even small chemical substitutions in organic molecules may significantly influence the hybrid crystal symmetry.[78] Finally, addressing the changes of the vibrational spectrum and phonon mode symmetries under electronic excitation[79,80] would constitute a crucial step towards understanding charge carrier-phonon coupling and polaron formation in MHPs. This can be addressed by future transient birefringence experiments with simultaneous optical injection of charge carriers.

Hidden, local symmetry breaking is potentially intrinsic and highly relevant to a wide range of nominally centrosymmetric materials, in which distinct local subunits may exhibit short-range inversion symmetry breaking, enabling functionalities forbidden by the global, long-range symmetry.[81,82] Local nano- and mesoscale symmetry breaking has been suggested, among others, in inorganic superconducting perovskites,[83] cuprates[84] or metal oxides like $VO_2$.[85] Our study demonstrates that analyzing the resonant ionic contributions to the nonlinear susceptibility provides a robust optical tool to investigate symmetry properties of such compounds.

## 4. Conclusion

In conclusion, we demonstrate selective, THz-driven coherent lattice control in layered PEA+-based RPP compounds with a varying degree of two-dimensional confinement. We identify a vibrational mode specific to the 2D system, which likely mediates the interactions between the inorganic 2D layers and organic spacer molecules. By a systematic polarization study of the THz-induced transient birefringence, we observe signatures of elusive, globally hidden inversion symmetry breaking, solely experienced by the lowest frequency inorganic cage modes. Coherent control of these vibrational modes could provide a promising handle for influencing RPP optoelectronic and spintronic behavior at ultrafast timescales. Further, spatially-resolved investigations of the microscopic nature of the hidden symmetry breaking may deliver crucial insights for harnessing emergent properties originating from the material's local or transient non-centrosymmetric character.

## 5. Experimental Section/Methods

*Crystal synthesis*

The single crystal samples were grown by the Anti-solvent Vapor-Assisted Crystallization AVC (free-standing crystals for n=1,2,3, Figure S1 a-c) or Anti-solvent Vapor-Assisted Crystallization AVCC (thin crystals on glass substrate for n=1,2, Fig. S1 d-e) methods. The details of the synthesis are described in Ref.[86]. The regular shape of the crystals allowed us to orient them relative to the field polarization directions. In Fig. S1, the direction corresponding to the long flake edge, for which $\psi = 0°$ when the axis is parallel to the THz field polarization direction, is marked.

*THz-induced transient birefringence measurements*

Single-cycle THz pump pulses (linearly polarized, peak fields ~1 MV/cm, 0.5-3 THz, $\tau_{\text{FWHM}} \approx 190$ fs) are used to induce the transient birefringence-are generated by optical rectification in $LiNbO_3$.[87] The $LiNbO_3$ crystal is driven by optical/NIR pulses (wavelength 800 nm, pulse duration 35 fs, pulse energy 5 mJ, repetition rate 1 kHz) provided by a Ti:Sapphire amplifier (Coherent Legend Elite Duo) in the tilted pulse front geometry.[87] Broadband pulses at 80 MHz repetition rate (wavelength 800 nm, 400 pJ, 20 fs pulse duration, linearly polarized) output from a Ti:Sapphire oscillator (Vitara) are used for probing. The THz pump (focus FWHM ~300 μm) and probe (focus FWHM ~50 μm) pulses are collinearly focused on the sample using a parabolic mirror and lens, respectively. A linear translation stage is used to control the temporal delay between the pump and probe pulses. For detection of the fields radiated by the nonlinear polarization, a heterodyne detection scheme is employed, with the residual transmitted probe acting as a local oscillator. The transmitted probe and nonlinear signal pass through a balancing setup consisting of a quarter-wave plate and half-wave plate and a Wollaston prism. The Wollaston prism projects the perpendicular vertical and horizontal polarization components of the local oscillator and signal fields onto two separate photodiodes and the differential signal is detected. The polarization of the probe is controlled by using a half-wave plate and is set to 45° relative to the fixed vertical THz polarization direction for all of the measurements, apart from the probe polarization scans described in the **Supporting Information, S12**. The room temperature measurements are performed with the sample mounted on a rotating holder inside of nitrogen purging box, and for low-temperature measurements, the sample is placed inside an optical, liquid nitrogen-cooled flow cryostat.

*Time-domain THz transmission spectroscopy*

THz pulses generated by optical rectification in $LiNbO_3$ (identical as in the transient birefringence measurements) were focused on the sample and the transmitted THz field was first collected by a second parabolic mirror, and then subsequently focused into an electro-optic sampling crystal (100 µm ZnTe), collinearly with an 800 nm probe beam. The same heterodyne detection scheme as described for the TKE experiments was used to record the transient birefringence.

*Raman spectroscopy*

Polarization-resolved static micro-Raman measurements were measured using the Horiba T64000 Raman triple grating spectrometer under ambient conditions with a 647 nm excitation laser in backscattering geometry. The incident beam was linearly polarized perpendicular to the 2D perovskite layer planes and the polarization direction relative to the sample controlled using a half waveplate. The scattered light was recorded by a spectrometer after passing through a linear polarization analyser, allowing to separately measure the co- and cross-polarized contributions.

**Acknowledgements:**

We thank Nimrod Benshalom, Alexander Fellows, Alexander Paarmann and Markus Raschke for fruitful discussions. Sebastian F. Maehrlein, Luca Perfetti and Emmanuelle Deleporte acknowledge funding of the 2D-HYPE project from Deutsche Forschungsgemeinschaft (German Research Foundation, grant No. 490867834) and the Agence Nationale de la Recherche (grant No. ANR-21-CE30-0059). Marie Cherasse acknowledges support of DAAD Scholarship 57507869 and Fondation L'Oréal-UNESCO with For Women in Science French Young Talents 2022 prize, in partnership with French Academy of Sciences.

**References**

[1]     L. Schmidt-Mende, V. Dyakonov, S. Olthof, F. Ünlü, K. M. T. Lê, S. Mathur, A. D. Karabanov, D. C.

Lupascu, L. M. Herz, A. Hinderhofer, F. Schreiber, A. Chernikov, D. A. Egger, O. Shargaieva, C. Cocchi, E. Unger, M. Saliba, M. M. Byranvand, M. Kroll, F. Nehm, K. Leo, A. Redinger, J. Höcker, T. Kirchartz, J. Warby, E. Gutierrez-Partida, D. Neher, M. Stolterfoht, U. Würfel, M. Unmüssig, J. Herterich, C. Baretzky, J. Mohanraj, M. Thelakkat, C. Maheu, W. Jaegermann, T. Mayer, J. Rieger, T. Fauster, D. Niesner, F. Yang, S. Albrecht, T. Riedl, A. Fakharuddin, M. Vasilopoulou, Y. Vaynzof, D. Moia, J. Maier, M. Franckevičius, V. Gulbinas, R. A. Kerner, L. Zhao, B. P. Rand, N. Glück, T. Bein, F. Matteocci, L. A. Castriotta, A. Di Carlo, M. Scheffler, C. Draxl, *APL Mater.* **2021**, *9*, 109202.

[2] L. Mao, C. C. Stoumpos, M. G. Kanatzidis, *J. Am. Chem. Soc.* **2019**, *141*, 1171.

[3] Y. Chen, Y. Sun, J. Peng, J. Tang, K. Zheng, Z. Liang, *Adv. Mater.* **2018**, *30*, 1703487.

[4] J. C. Blancon, A. V. Stier, H. Tsai, W. Nie, C. C. Stoumpos, B. Traoré, L. Pedesseau, M. Kepenekian, F. Katsutani, G. T. Noe, J. Kono, S. Tretiak, S. A. Crooker, C. Katan, M. G. Kanatzidis, J. J. Crochet, J. Even, A. D. Mohite, *Nat. Commun.* **2018**, *9*, 2254.

[5] Z. Guo, J. Wang, W. J. Yin, *Energy Environ. Sci.* **2022**, *15*, 660.

[6] M. Zacharias, G. Volonakis, F. Giustino, J. Even, *npj Comput. Mater.* **2023**, *9*, 153.

[7] M. Frenzel, M. Cherasse, J. M. Urban, F. Wang, B. Xiang, L. Nest, L. Huber, L. Perfetti, M. Wolf, T. Kampfrath, X. Y. Zhu, S. F. Maehrlein, *Sci. Adv.* **2023**, *9*, eadg3856.

[8] M. J. Schilcher, P. J. Robinson, D. J. Abramovitch, L. Z. Tan, A. M. Rappe, D. R. Reichman, D. A. Egger, *ACS Energy Lett.* **2021**, *6*, 2162.

[9] O. Yaffe, Y. Guo, L. Z. Tan, D. A. Egger, T. Hull, C. C. Stoumpos, F. Zheng, T. F. Heinz, L. Kronik, M. G. Kanatzidis, J. S. Owen, A. M. Rappe, M. A. Pimenta, L. E. Brus, *Phys. Rev. Lett.* **2017**, *118*, 136001.

[10] Z. Zhang, J. Zhang, Z.-J. Liu, N. S. Dahodl, W. Paritmongkol, N. Brown, Y.-C. Chien, Z. Dai, K. A. Nelson, W. A. Tisdale, A. M. Rappe, E. Baldini, **2023**, eadg4417.

[11] Y.-F. Chen, A. Mahata, A. D. Lubio, M. Cinquino, A. Coriolano, L. Skokan, Y.-G. Jeong, L. Razzari, L. De Marco, A. Ruediger, F. De Angelis, S. Colella, E. Orgiu, *Adv. Opt. Mater.* **2021**, *10*, 2100439.

[12] N. S. Dahod, A. France-Lanord, W. Paritmongkol, J. C. Grossman, W. A. Tisdale, *J. Chem. Phys.* **2020**, *153*, 044710.

[13] K. Lee, S. F. Maehrlein, X. Zhong, D. Meggiolaro, J. C. Russell, D. A. Reed, B. Choi, F. De Angelis, X. Roy, X. Zhu, *Adv. Mater.* **2019**, *31*, 1903209.

[14] H. Long, X. Peng, J. Lu, K. Lin, L. Xie, B. Zhang, L. Ying, Z. Wei, *Nanoscale* **2019**, *11*, 21867.

[15] Y. Lu, Q. Wang, L. Han, Y. Zhao, Z. He, W. Song, C. Song, Z. Miao, *Adv. Funct. Mater.* **2024**, *34*, 2314427.

[16] K. Leng, R. Li, S. P. Lau, K. P. Loh, *Trends Chem.* **2021**, *3*, 716.

[17] B. Maurer, C. Vorwerk, C. Draxl, *Phys. Rev. B* **2022**, *105*, 155149.

[18] M. Kepenekian, J. Even, *J. Phys. Chem. Lett.* **2017**, *8*, 3362.

[19] C.-K. Yang, W.-N. Chen, Y.-T. Ding, J. Wang, Y. Rao, W.-Q. Liao, Y.-Y. Tang, P.-F. Li, Z.-X. Wang, R.-G. Xiong, *Adv. Mater.* **2019**, *31*, 180808.

[20] D. J. Morrow, M. P. Hautzinger, D. P. Lafayette, J. M. Scheeler, L. Dang, M. Leng, D. D. Kohler, A. M. Wheaton, Y. Fu, I. A. Guzei, J. Tang, S. Jin, J. C. Wright, *J. Phys. Chem. Lett.* **2020**, *11*, 6551.

[21] K. Frohna, T. Deshpande, J. Harter, W. Peng, B. A. Barker, J. B. Neaton, S. G. Louie, O. M. Bakr, D. Hsieh, M. Bernardi, *Nat. Commun.* **2018**, *9*, 1829.

[22] J. Yin, P. Maity, L. Xu, A. M. El-Zohry, H. Li, O. M. Bakr, J. L. Brédas, O. F. Mohammed, *Chem. Mater.* **2018**, *30*, 8538.

[23] E. Lafalce, R. Bodin, B. W. Larson, J. Hao, M. A. Haque, U. Huynh, J. L. Blackburn, Z. V. Vardeny, *ACS Nano* **2024**, *18*, 18299.

[24] Y. Zhai, S. Baniya, C. Zhang, J. Li, P. Haney, C. X. Sheng, E. Ehrenfreund, Z. V. Vardeny, *Sci. Adv.* **2017**, *3*, e1700704.

[25] S. B. Todd, D. B. Riley, A. Binai-Motlagh, C. Clegg, A. Ramachandran, S. A. March, J. M. Hoffman, I. G. Hill, C. C. Stoumpos, M. G. Kanatzidis, Z. G. Yu, K. C. Hall, *APL Mater.* **2019**, *7*, 081116.

[26] H. Lu, Z. V. Vardeny, M. C. Beard, *Nat. Rev. Chem.* **2022**, *6*, 470.

[27] W. Mihalyi-koch, Z. Dai, X. Lu, A. M. Rappe, S. Jin, W. Mihalyi-koch, S. Guo, Z. Dai, D. Pan, D. P. L. li, J. M. Scheeler, K. M. Sanders, S. J. Teat, J. C. Wright, X. Lu, A. M. Rappe, *CHEMPR* **2024**, *10*, 2180.

[28] M. K. Jana, R. Song, Y. Xie, R. Zhao, P. C. Sercel, V. Blum, D. B. Mitzi, *Nat. Commun.* **2021**, *12*, 4982.

[29] S. Zuri, A. Shapiro, L. Kronik, E. Lifshitz, *J. Phys. Chem. Lett.* **2023**, *14*, 4901.

[30] E. Lafalce, E. Amerling, Z. G. Yu, P. C. Sercel, L. Whittaker-Brooks, Z. V. Vardeny, *Nat. Commun.* **2022**, *13*, 483.

[31] P. P. Joshi, S. F. Maehrlein, X. Zhu, *Adv. Mater.* **2019**, *31*, 180305.

[32] Y. Yamada, Y. Kanemitsu, *NPG Asia Mater.* **2022**, *14*, 48.

[33] T. Etienne, E. Mosconi, F. De Angelis, *J. Phys. Chem. Lett.* **2016**, *7*, 1638.

[34] A. N. Beecher, O. E. Semonin, J. M. Skelton, J. M. Frost, M. W. Terban, H. Zhai, A. Alatas, J. S. Owen, A. Walsh, S. J. L. Billinge, *ACS Energy Lett.* **2016**, *1*, 880.

[35] J. Fu, T. Bian, J. Yin, M. Feng, Q. Xu, Y. Wang, T. C. Sum, *Nat. Commun.* **2024**, *15*, 4562.

[36] T. Huber, M. Ranke, A. Ferrer, L. Huber, S. L. Johnson, *Appl. Phys. Lett.* **2015**, *107*, 091107.

[37] M. C. Hoffmann, N. C. Brandt, H. Y. Hwang, K. Lo Yeh, K. A. Nelson, *Appl. Phys. Lett.* **2009**, *95*, 23110.

[38] Z. Chen, Q. Zhang, M. Zhu, H. Chen, X. Wang, S. Xiao, K. P. Loh, G. Eda, J. Meng, J. He, *J. Phys. Chem. Lett.* **2021**, *12*, 7010.

[39] I. Abdelwahab, P. Dichtl, G. Grinblat, K. Leng, X. Chi, I. H. Park, M. P. Nielsen, R. F. Oulton, K. P. Loh, S. A. Maier, *Adv. Mater.* **2019**, *31*, 1902685.

[40] W. Liu, J. Xing, J. Zhao, X. Wen, K. Wang, P. Lu, Q. Xiong, *Adv. Opt. Mater.* **2017**, *5*, 1601045.

[41] S. Maehrlein, A. Paarmann, M. Wolf, T. Kampfrath, *Phys. Rev. Lett.* **2017**, *119*, 127402.

[42] M. Menahem, Z. Dai, S. Aharon, R. Sharma, M. Asher, Y. Diskin-Posner, R. Korobko, A. M. Rappe, O. Yaffe, *ACS Nano* **2021**, *15*, 10153.

[43] L. Huber, S. F. Maehrlein, F. Wang, Y. Liu, X. Y. Zhu, *J. Chem. Phys.* **2021**, *154*, 094202.

[44] M. L. Lin, B. Dhanabalan, G. Biffi, Y. C. Leng, S. Kutkan, M. P. Arciniegas, P. H. Tan, R. Krahne, *Small* **2022**, *18*, 2106759.

[45] M. Mączka, M. Ptak, *Solids* **2022**, *3*, 111.

[46] B. Dhanabalan, Y. C. Leng, G. Biffi, M. L. Lin, P. H. Tan, I. Infante, L. Manna, M. P. Arciniegas, R. Krahne, *ACS Nano* **2020**, *14*, 4689.

[47] V. A. Dragomir, S. Neutzner, C. Quarti, D. Cortecchia, A. Petrozza, S. Roorda, D. Beljonne, R. Leonelli, A. R. S. Kandada, C. Silva, *arXiv:*1812.05255 [cond-mat.mtrl-sci], submitted: Dec **2018**.

[48] E. Y. Chen, B. Monserrat, *J. Phys. Chem. C* **2024**, *128*, 12194.

[49] R.-I. Biega, M. Bokdam, K. Herrmann, J. Mohanraj, D. Skyrbeck, M. Thelakkat, M. Retsch, L. Leppert, **2023**, *127*, 9183.

[50] J. Fu, M. Li, A. Solanki, Q. Xu, Y. Lekina, S. Ramesh, Z. X. Shen, T. C. Sum, *Adv. Mater.* **2021**, *33*, 1.

[51] P. Guo, Y. Xia, J. Gong, D. H. Cao, X. Li, X. Li, Q. Zhang, C. C. Stoumpos, M. S. Kirschner, H. Wen, V. B. Prakapenka, J. B. Ketterson, A. B. F. Martinson, T. Xu, M. G. Kanatzidis, M. K. Y. Chan, R. D. Schaller, *Adv. Funct. Mater.* **2020**, *30*, 1907982.

[52] Z. Liu, Y. Shi, T. Jiang, L. Luo, C. Huang, M. Mootz, Z. Song, Y. Yan, Y. Yao, J. Zhao, J. Wang, *PRX Energy* **2024**, *3*, 023009.

[53] L. N. Quan, Y. Park, P. Guo, M. Gao, J. Jin, J. Huang, J. K. Copper, A. Schwartzberg, R. Schaller, D. T. Limmer, P. Yang, *Pnas* **2021**, *118*, e210442511.

[54] X. Gao, Y. Wu, Y. Zhang, X. Chen, Z. Song, T. Zhang, Q. Fang, Q. Ji, M. G. Ju, J. Wang, *Small Methods* **2024**, *2401172*.

[55] M. Nagai, T. Tomioka, M. Ashida, M. Hoyano, R. Akashi, Y. Yamada, T. Aharen, Y. Kanemitsu, *Phys. Rev. Lett.* **2018**, *121*, 145506.

[56] M. Balkanski, R. F. Wallis, E. Haro, *Phys. Rev. B* **1983**, *28*, 1928.

[57] P. G. Klemens, *Phys. Rev. B* **1975**, *11*, 3206.

[58] M. Hase, K. Ishioka, M. Kitajima, K. Ushida, S. Hishita, *Appl. Phys. Lett.* **2000**, *76*, 1258.

[59] H. Yang, S. Mandal, B. Li, T. K. Ghosh, J. M. Peterson, P. Guo, L. Dou, M. Chen, L. Huang, **2024**, *146*, 33928.

[60] L. Lodeiro, X. Liang, A. Walsh, *J. Phys. Chem. C* **2024**, *128*, 20947.

[61] D. M. Juraschek, S. F. Maehrlein, *Phys. Rev. B* **2018**, *97*, 174302.

[62] M. Fox, *Optical Properties of Solids*, 2nd ed., Oxford University Press, **2010**.

[63] Deepak Thrithamarassery Gangadharan, Y. Han, A. Dubey, X. Gao, B. Sun, Q. Qiao, R. Izquierdo, D. Ma, *Sol. RRL* **2018**, *2*, 1700215.

[64] D. Seyitliyev, X. Qin, M. K. Jana, S. M. Janke, X. Zhong, W. You, D. B. Mitzi, V. Blum, K. Gundogdu, *Adv. Funct. Mater.* **2023**, *33*, 2213021.

[65] J. J. Geuchies, J. Klarbring, L. Di Virgillio, S. Fu, S. Qu, G. Liu, H. Wang, J. M. Frost, A. Walsh, M. Bonn, H. Kim, *Nano Lett.* **2024**, *28*, 8642.

[66] E. J. Ayars, H. D. Hallen, C. L. Jahncke, *Phys. Rev. Lett.* **2000**, *85*, 4180.

[67] A. Ashoka, S. Nagane, N. Strkalj, B. Roose, J. Sung, J. L. Macmanus-Driscoll, S. D. Stranks, S. Feldmann, A. Rao, *Nat. Mater.* **2023**, *22*, 977.

[68] S. Zuri, L. Kronik, E. Lifshitz, *J. Phys. Chem. Lett.* **2024**, 11637.

[69] M. Telychko, S. Edalatmanesh, K. Leng, I. Abdelwahab, N. Guo, C. Zhang, J. I. Mendieta-Moreno, M. Nachtigall, J. Li, K. P. Loh, P. Jelínek, J. Lu, *Sci. Adv.* **2022**, *8*, eabj0395.

[70] D. R. Kripalani, Q. Guan, H. Yan, Y. Cai, K. Zhou, *ACS Nano* **2024**, *18*, 14187.

[71] A. N. Beecher, O. E. Semonin, J. M. Skelton, J. M. Frost, M. W. Terban, H. Zhai, A. Alatas, J. S. Owen, A. Walsh, S. J. L. Billinge, *ACS Energy Lett.* **2016**, *1*, 880.

[72] S. McKechnie, J. M. Frost, D. Pashov, P. Azarhoosh, A. Walsh, M. Van Schilfgaarde, *Phys. Rev. B* **2018**, *98*, 085108.

[73] V. Balos, T. Garling, A. D. Duque, B. John, M. Wolf, M. Thämer, *J. Phys. Chem. C* **2022**, *126*, 10818.

[74] X. Li, J. Li, Y. Li, A. Ozcan, M. Jarrahi, *Light Sci. Appl.* **2023**, *12*, 233.

[75] H. T. Stinson, A. Sternbach, O. Najera, R. Jing, A. S. Mcleod, T. V. Slusar, A. Mueller, L. Anderegg, H. T. Kim, M. Rozenberg, D. N. Basov, *Nat. Commun.* **2018**, *9*, 3604.

[76] Y. Liu, L. Collins, R. Proksch, S. Kim, B. R. Watson, B. Doughty, T. R. Calhoun, M. Ahmadi, A. V. Ievlev, S. Jesse, S. T. Retterer, A. Belianinov, K. Xiao, J. Huang, B. G. Sumpter, S. V. Kalinin, B. Hu, O. S. Ovchinnikova, *Nat. Mater.* **2018**, *17*, 1013.

[77] J. Hu, I. W. H. Oswald, S. J. Stuard, M. M. Nahid, N. Zhou, O. F. Williams, Z. Guo, L. Yan, H. Hu, Z. Chen, X. Xiao, Y. Lin, Z. Yang, J. Huang, A. M. Moran, H. Ade, J. R. Neilson, W. You, *Nat. Commun.* **2019**, *10*, 1276.

[78] T. Schmitt, S. Bourelle, N. Tye, G. Soavi, A. D. Bond, S. Feldmann, B. Traore, C. Katan, J. Even, S. E. Dutton, F. Deschler, *J. Am. Chem. Soc.* **2020**, *142*, 5060.

[79] M. Dyksik, D. Beret, M. Baranowski, H. Duim, S. Moyano, K. Posmyk, A. Mlayah, S. Adjokatse, D. K. Maude, M. A. Loi, P. Puech, P. Plochocka, *Adv. Sci.* **2024**, *11*, 2305182.

[80] L. Ni, U. Huynh, A. Cheminal, T. H. Thomas, R. Shivanna, T. F. Hinrichsen, S. Ahmad, A. Sadhanala, A. Rao, *ACS Nano* **2017**, *11*, 10834.

[81] L. Yuan, Q. Liu, X. Zhang, J. W. Luo, S. S. Li, A. Zunger, *Nat. Commun.* **2019**, *10*, 906.

[82] Q. Liu, X. Zhang, H. Jin, K. Lam, J. Im, A. J. Freeman, A. Zunger, *Phys. Rev. B - Condens. Matter Mater. Phys.* **2015**, *91*, 235204.

[83] S. Griffitt, M. Spaić, J. Joe, Z. W. Anderson, D. Zhai, M. J. Krogstad, R. Osborn, D. Pelc, M. Greven, *Nat. Commun.* **2023**, *14*, 845.

[84] K. Gotlieb, C.-Y. Lin, M. Serbyn, W. Zhang, C. L. Smallwood, C. Jozwiak, H. Eisaki, Z. Hussain, A. Vishwanath, A. Lanzara, *Science (80-. ).* **2018**, *362*, 1271.

[85] W. Liang, Z. Zhuo, Y. Ji, C. Lu, M. Gao, H. Yang, C. Chen, F. Pan, Y. Lin, *npj Quantum Mater.* **2019**, *4*, 39.

[86] F. Lédée, G. Trippé-Allard, H. Diab, P. Audebert, D. Garrot, J. S. Lauret, E. Deleporte, *CrystEngComm* **2017**, *19*, 2598.

[87] H. Hirori, A. Doi, F. Blanchard, K. Tanaka, *Appl. Phys. Lett.* **2011**, *98*, 091106.

# Supporting Information

## for

## "THz-Driven Coherent Phonon Fingerprints of Hidden Symmetry Breaking in 2D Layered Hybrid Perovskites"

Joanna M. Urban[1]*, Michael S. Spencer[1], Maximilian Frenzel[1], Gaëlle Trippé- Allard[2], Marie Cherasse[1,3]‡, Charlotte Berrezueta Palacios[4], Olga Minakova[1], Eduardo Bedê Barros[5,6], Luca Perfetti[3], Stephanie Reich[4], Martin Wolf[1], Emmanuelle Deleporte[2], Sebastian F. Maehrlein[1,7,8]*

[1] *Fritz Haber Institute of the Max Planck Society, Department of Physical Chemistry, Faradayweg 4-6, 14195 Berlin, Germany*

[2] *Lumière, Matière et Interfaces (LuMIn) Laboratory, Université Paris-Saclay, ENS Paris-Saclay, CentraleSupélec, CNRS, 91190 Gif-sur-Yvette, France*

[3] *Laboratoire des Solides Irradiés, CEA/DRF/IRAMIS, École Polytechnique, CNRS, Institut Polytechnique de Paris, F-91128 Palaiseau, France*

[4] *Department of Physics, Freie Universität Berlin, Berlin 14195, Germany*

[5] *Department of Physics, Universidade Federal do Ceara, Fortaleza, Ceara, 60455-760 Brazil*

[6] *Institut für Festkörperphysik, Technische Universität Berlin, Hardenbergstraße 36, 10623 Berlin, Germany*

[7] *Helmholtz-Zentrum Dresden-Rossendorf, Institute of Radiation Physics, Dresden, Germany*

[8] *Technische Universität Dresden, Institute of Applied Physics, Dresden, Germany*

‡ Currently at Laboratoire d'Optique Appliquée, ENSTA Paris, CNRS, École Polytechnique, Institut Polytechnique de Paris, 91761 Palaiseau, France.

*email: maehrlein@fhi-berlin.mpg.de, urban@fhi-berlin.mpg.de

**Section S1. Sample growth and morphology**

The samples were grown as free-standing crystals by the Anti-solvent Vapor-Assisted Crystallization (AVC) method as well as, for n=1,2, by the Anti-solvent Vapor-Assisted Crystallization (AVCC) method on BK7 glass substrates. The AVCC crystals were measured as-grown on the substrate (**Figure S1 d-e**)) and the AVC crystal either as-grown (**Figure S1 a-c**) or after thinning down by scotch tape cleaving. The regular shape of the crystals allowed us to orient them relative to the field polarization directions. In Figure S1 d, e, the direction corresponding to the long axis of the AVCC crystals, defining the azimuthal angle $\psi = 0°$ when parallel to the THz field polarization direction, is marked. For the AVC crystals, the short/long edge assignment was more ambiguous but it was possible to distinguish between directions parallel and diagonal to the crystal edges.

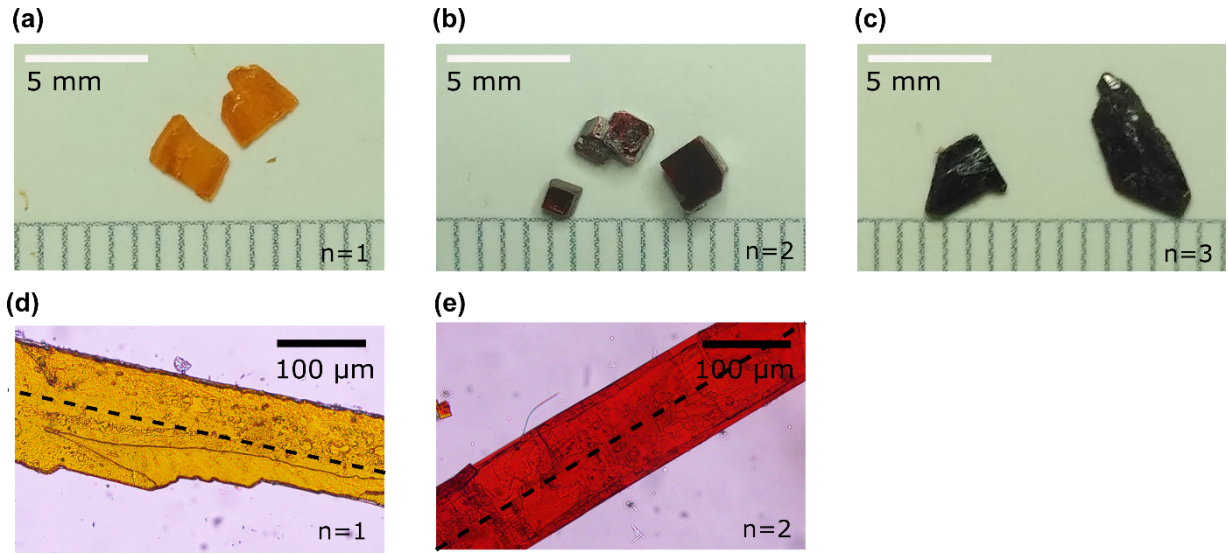

**Figure S1.** Optical images of **a)** AVC $n$=1, **b)** AVC $n$=2, **c)** AVC $n$=3, **d)** AVCC $n$=1, **e)** AVCC $n$=2 crystals. The dashed line marks the flake long axis.

**Section S2. Influence of sample thickness on transient birefringence signal**

To exclude possible misinterpretation of the oscillatory transient birefringence signals due to propagation effects,[1,2] measurements for the $n$=1 compound were performed on multiple samples with different thicknesses. **Figure S2** shows a comparison of the room temperature signals measured on a very thin ($d < 1\ \mu$m) AVCC crystal on BK7 substrate and on an as-grown free-standing AVC crystal of $d \sim 230\ \mu$m thickness. Features related to propagation artifacts should be strongly thickness dependent, so the observation of very similar peaks in the spectra of the two different samples in **Figure S2b** (0.8-0.95 THz, 1.15-1.35 THz and 1.55-1.65 THz) confirms their assignment as coherent phonon signatures. Especially the ~0.9 THz mode clearly dominates the signals for longer cutoff times (**Figure S2c**) for both samples. A smearing of the instantaneous response features in time domain is visible for the thicker sample in the form of a slight exponential tail.[1] Slight differences between the spectra of the two samples observed at shorter cut-off times (**Figure S2b**) may be related to interferences of the lattice response with residual contributions from the temporally smeared out instantaneous nonresonant response in the case of the thicker sample.

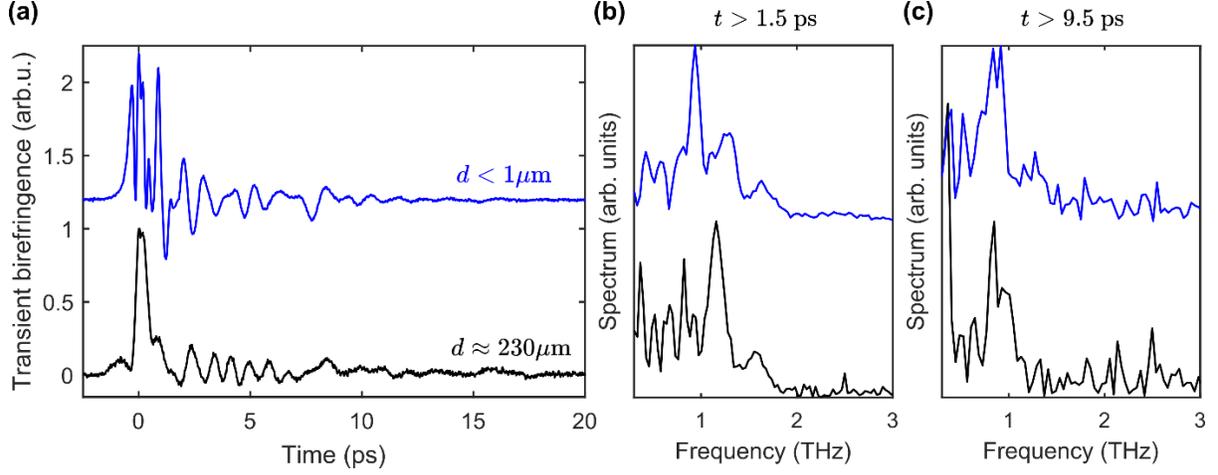

**Figure S2**. **a)** Transient birefringence signal recorded for a thin AVCC (blue) and thick AVC (black) crystals at $\psi = 0°$. **b)** Corresponding FTs of the oscillatory signal for cut-off time 1.5 ps and **c)** 9.5 ps.

**Section S3. Sample thickness: pump-probe walk-off effects**

We note differences between the relative intensities of the different phonon signatures in the signals measured on different samples, with the higher frequency modes typically becoming less prominent in thicker AVC samples compared to thinner AVCC crystals. **Figure S3a** compares the FTs of transient birefringence signals measured on four different samples of n=1, thin AVCC (typically $d < 5\,\mu$m), a thick AVCC sample and an AVC sample (typically $d > 50\,\mu$m). The walk-off effects due to different propagation velocity of the pump and probe pulse[3] will lead to a relative decrease of the higher-frequency components of the driving force at finite depths in the sample, as demonstrated by simple modelling in **Figure S3 b, c**. For simplicity, we consider only the real part of the refractive index, neglecting absorption and anisotropy, and assume $n_{\mathrm{THz}} = 2.9$ (calculated from approximate $\tilde{\epsilon}_\infty$ based on data from ref.[4]) and $n_{800} = 2$ (extrapolating after[5]). The incident field is extracted from an electro-optic sampling measurement after detector response function correction.[6] The driving force spectrum for the photonic driving mechanism is proportional to the Fourier transform of the squared THz field:

$$F_{\mathrm{dr}}(\omega) \propto \mathrm{FT}\left(E_{\mathrm{THz}}^2(t)\right) \tag{S1}$$

The THz-induced Kerr-type transient birefringence signal $S(t)$ related to instantaneous electronic polarizability can be written as being proportional to:[7]

$$S(t) \propto \int_0^d dz\, \Delta n\left(z, t + \frac{z}{v_{800}}\right) = \int_0^d dz\, \Delta n(0, t + \beta z) \propto \int_0^d dz\, F_{\mathrm{dr}}(0, t + \beta z) \tag{S2}$$

where $\beta = v_{800}^{-1} - v_{\text{THz}}^{-1} = (n_{800} - n_{\text{THz}})c_0$ is the inverse velocity mismatch between the probe and the pump and $\Delta n(z,t) = n_2 c_0 \varepsilon_0 E_{\text{THz}}^2(z,t) \propto F_{\text{dr}}(z,t)$ is the THz-induced refractive index anisotropy determined by the square of the THz electric field and the nonlinear refractive index $n_2$, assuming the THz field after propagating a distance $z$ inside the sample can be written as $E_{\text{THz}}(z,t) = E_{\text{THz}}(z=0, t - zv_{\text{THz}}^{-1})$[7] and neglecting THz absorption. In frequency domain:

$$S(\omega) \propto \text{FT}\left(\frac{1}{\beta} F_{\text{dr}}(0,t) * \text{rect}\left(\frac{t}{\beta d} - \frac{1}{2}\right)\right)$$

$$\propto \text{FT}\left(E_{\text{THz}}^2(0,t)\right) \cdot d \cdot \exp\left(-\frac{i\omega\beta d}{2}\right) \text{sinc}\left(\frac{\omega\beta d}{2}\right) = F_{\text{eff,dr}}(\omega, d) \qquad (S3)$$

where the last expression describes a modified effective driving force spectrum $F_{\text{eff,dr}}(\omega, d)$ resulting from integration over the sample thickness. Figure S3b shows the effective driving force spectrum as a function of the sample thickness compared to the driving force calculated for the field at $z=0$. In frequency domain (Figure S3b), the driving force spectral amplitude bandwidth decreases significantly with increasing sample thickness. In time domain (Figure S3c), the velocity mismatch leads to a smearing out of the electronic response signal and appearance of box-like signals for larger thicknesses.[7] Including anisotropy, the dispersion of the THz refractive index and absorption which are both significant in (PEA)$_2$PbI$_4$ in the 0-5 THz range (see Section S4) to fully describe our system would result in a significantly more complex response, but maintaining a similar driving force bandwidth reduction for thicker samples.

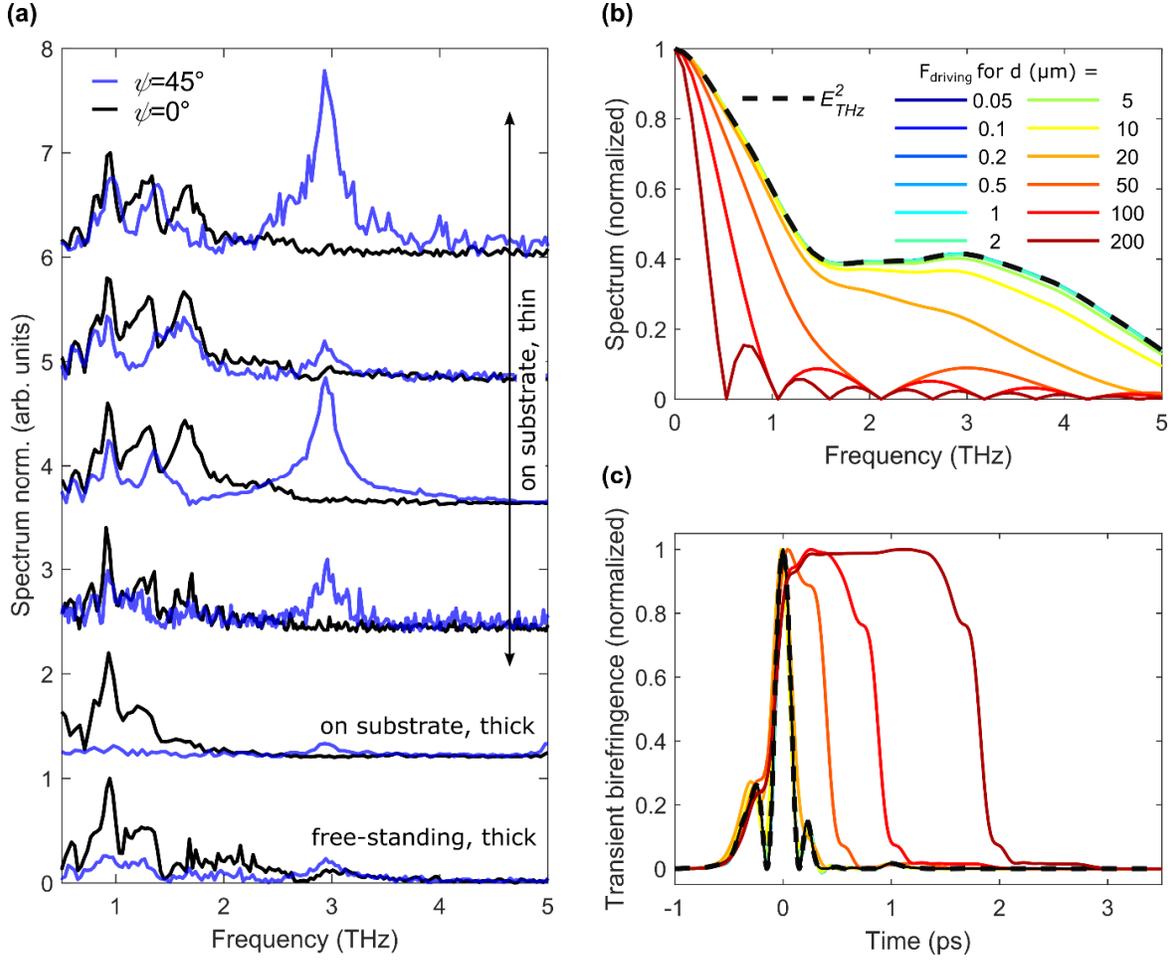

**Figure S3 a)** Fourier transforms of transient birefringence spectra for sample azimuthal orientations $\psi = 0°$ and $\psi = 45°$ for thin and thick AVCC crystals and thick AVC free-standing crystals. Each pair of spectra for the two angles for each sample were normalized by the intensity of the 0.9 THz peak for $\psi = 0°$. **b)** Driving force spectrum calculated based on the incident field (black, dashed) and considering walk-off effects for a range of thicknesses $d$. **c)** Corresponding time domain TKE instantaneous response signals.

### Section S4. Absorption and dispersion effects

**Figure S4a** shows the complex refractive index terms for $(PEA)_2PbI_4$ calculated from the dielectric constant given in Ref.[4]. The frequency-dependent penetration depth of the THz into the sample can be estimated based on the absorption coefficient as (**Figure S4b**):

$$l = \alpha^{-1} = \frac{\lambda_0}{4\pi k} \quad (S4)$$

**Figure S4c** shows the modification of the spectrum of the THz field and corresponding driving force upon transmission into the sample taking into account the transmission coefficient at the air-sample interface:

$$t = \frac{2}{(\tilde{n}_{THz} + 1)} \quad (S5)$$

**Figure S4d** shows the spectra of the THz electric field and the driving force after propagation for different finite depths into the sample. Higher frequencies are more strongly absorbed, leading to a reduction of the field bandwidth for larger thicknesses. Together with propagation effects, absorption could explain the higher contribution from higher frequency modes in thinner samples.

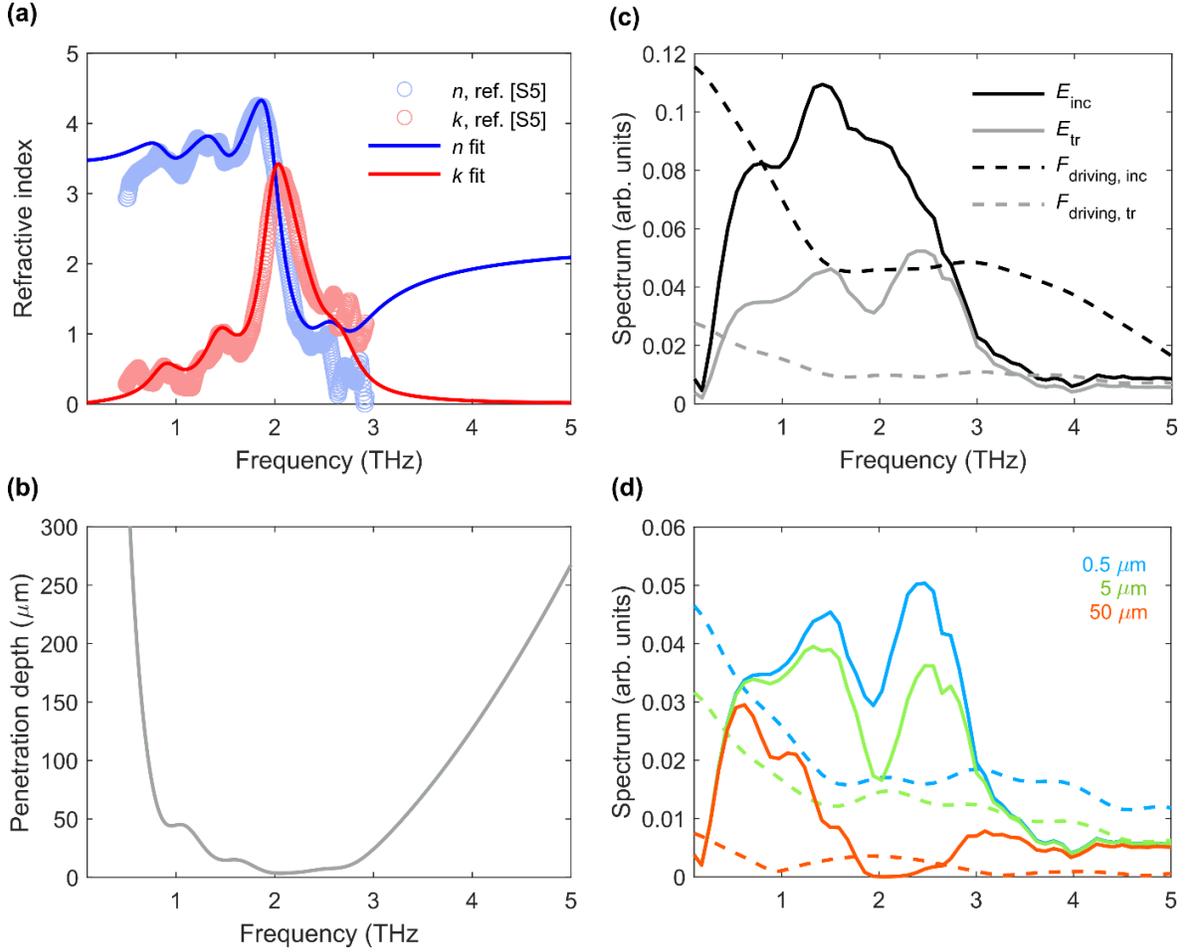

**Figure S4 a)** Complex refractive index $\tilde{n} = n + ik$ calculated based on the dielectric function from ref. [4] and its fits using three Lorentzian oscillators. **b)** Penetration depth in the THz range calculated based on *k*. **c)** The incident THz electric field spectrum $E_{\mathrm{inc}}$, the field transmitted through the air-sample interface $E_{\mathrm{tr}}$ and the respective nonlinear driving force spectra. **d)** Electric field spectra and driving force spectra at a depth $z =$ 0.5, 5 and 50 µm in the sample.

### Section S5. Estimation of nonlinear refractive index

The nonlinear refractive index at ~1 THz is estimated for the *n*=1 sample by converting the measured transient birefringence signal into a rotation angle (**Figure S5**) and extracting the instantaneous peak rotation $\Delta\phi_{max} = 31.1$ mrad.

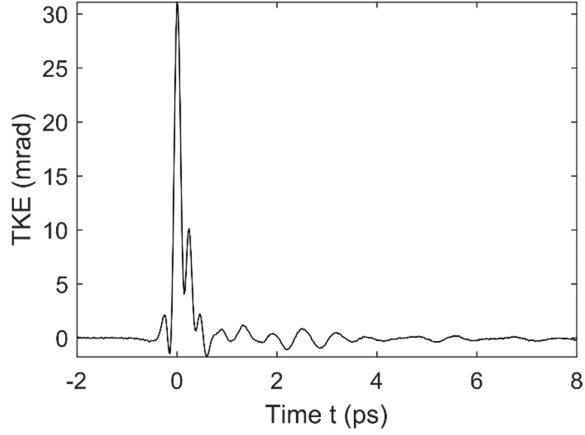

**Figure S5** Transient birefringence signal converted into rotation angle for n=1 AVCC sample at $\psi = 0°$.

Assuming $n_{1\text{THz}} = 3.51$ and $k_{1\text{THz}} = 0.53$ (see Figure S4a), a thickness of the sample $d = 5\mu m$, an absorption coefficient $\alpha_{1\text{THz}} = 4\pi k_{1\text{THz}} \lambda_{1\text{THz}}^{-1} = 2235.5$ cm$^{-1}$, an incident THz peak electric field $E_{peak} = 1.1$ MV/cm and an air-sample transmission coefficient given by Eq. (S5) $t_{1\text{THz}}$, we can write a ballpark estimate for $n_2$:[1,7]

$$n_2(1\text{ THz}) = \frac{\Delta\phi_{max} \cdot \lambda_{800}}{2\pi c_0 \varepsilon_0 \int_0^d dz (E_{1\text{THz}}(z))^2} =$$
$$= \frac{\Delta\phi_{max} \cdot \lambda_{800}}{2\pi c_0 \varepsilon_0 \int_0^d dz \, (t_{1\text{THz}} \cdot E_{peak} \cdot \exp(-0.5 \cdot \alpha_{1\text{THz}} \cdot z))^2} \tag{S6}$$

which gives a value of $n_2(1\text{THz}) \approx 1.4 \cdot 10^{-12}$ cm/W. The main source of uncertainty in the estimate of $n_2$ is the unknown exact sample thickness. The typical thickness of AVCC crystals is in the 0.5-10 µm range. Assuming 0.5 µm and 10 µm thicknesses for the calculations yields upper and lower limits for the estimated values of $n_2(1\text{THz}) \approx 1.3 \cdot 10^{-11}$ and $n_2(1\text{THz}) \approx 7.2 \cdot 10^{-13}$, respectively.

**Section S6. Probe fluence dependence**

We measured the transient birefringence signals as a function of varying probe fluence for the *n*=1 sample at two different orientations (long flake axis at $\psi = 0°$ and $\psi = 45°$ to THz polarization) to confirm that there are no additional effects caused by the presence of charge carriers generated by two-photon absorption and that the signal scales according to our predictions. The normalized signals (**Figure S5a** and **b**) and the corresponding FTs (**Figure S5c** and **d**) do not show any changes with the probe fluence. In the four wave mixing picture, the nonlinear signal field $E_{\text{signal}}^{(3)}$ is proportional to the probe electric field at the depth $z$ $E_{\text{probe}}(z)$ in the sample where the signal is generated. Assuming that most of the signal is generated near the surface of incidence, we can assume $E_{\text{probe}}(z) \approx E_{\text{probe,incident}} \sim \sqrt{I_{\text{probe}}}$ where $I_{\text{probe}}$ is the measured power of the incident probe. The signal measured in our detection scheme assuming balanced conditions is proportional to $(E^{(3)}_{\text{signal,x}} - E^{(3)}_{\text{signal,y}}) \cdot E_{\text{probe,tr}}$, where $E_{\text{probe,tr}}$ is the transmitted probe field after the sample used for heterodyning. The transmitted field can experience losses due to scattering and two-photon absorption, therefore is not simply proportional to $E_{\text{probe,incident}}$. We measure the photodiode voltage $V_{\text{PD}}$ on a single channel under balanced conditions for every $I_{\text{probe}}$, assuming that $E_{\text{probe,tr}} \sim V_{\text{PD}}$. We finally plot the transient birefringence signal magnitude as a function of $\sqrt{I_{\text{probe}} \cdot V_{\text{PD}}}$, and obtain almost perfect linear scaling as expected (**Figure S5e** and **f**).

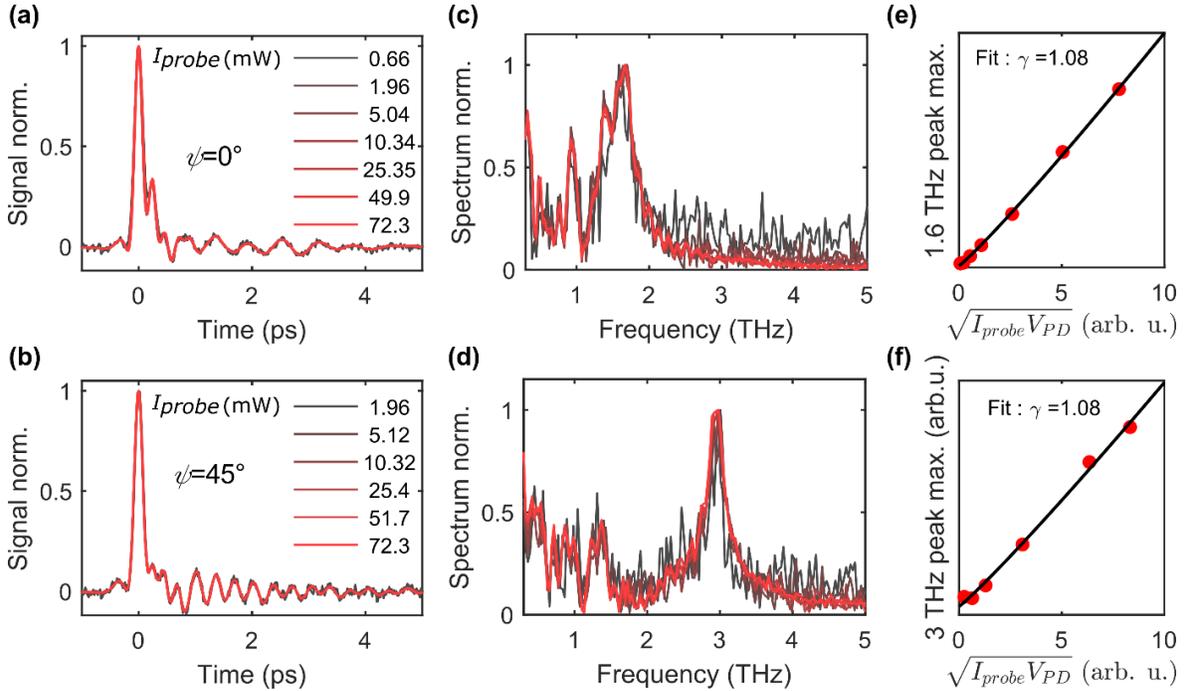

**Figure S6**. Normalized transient birefringence signals as a function of incident probe power for **a)** $\psi = 0°$ and **b)** $\psi = 45°$ and corresponding FTs **c)** and **d)**. Amplitude of the **e)** 1.6 THz peak and **f)** 3 THz peak as a function of $\sqrt{I_{\text{probe}} \cdot V_{\text{PD}}}$ and the fitted power laws $A = (\sqrt{I_{\text{probe}} \cdot V_{\text{PD}}})^\gamma$.

### Section S7. Reproducibility of azimuthal angle scans

We repeated the transient birefringence measurement as a function of the crystal azimuthal angle on several flakes to confirm the observed polar patterns. **Figure S7** shows a comparison of the angle-dependent instantaneous response amplitude and the Raman-active mode signal amplitudes for two different flakes in **a** and **b**. In both data sets, we observe deviations of the experimental signal from a four-fold symmetry for the instantaneous response as well as from the theoretically predicted coherent phonon signal amplitude. The differing deviations for the two different flakes suggest that these are experimental artifacts arising due to sample inhomogeneities and/or imperfect sample orientation in the plane normal to the beam.

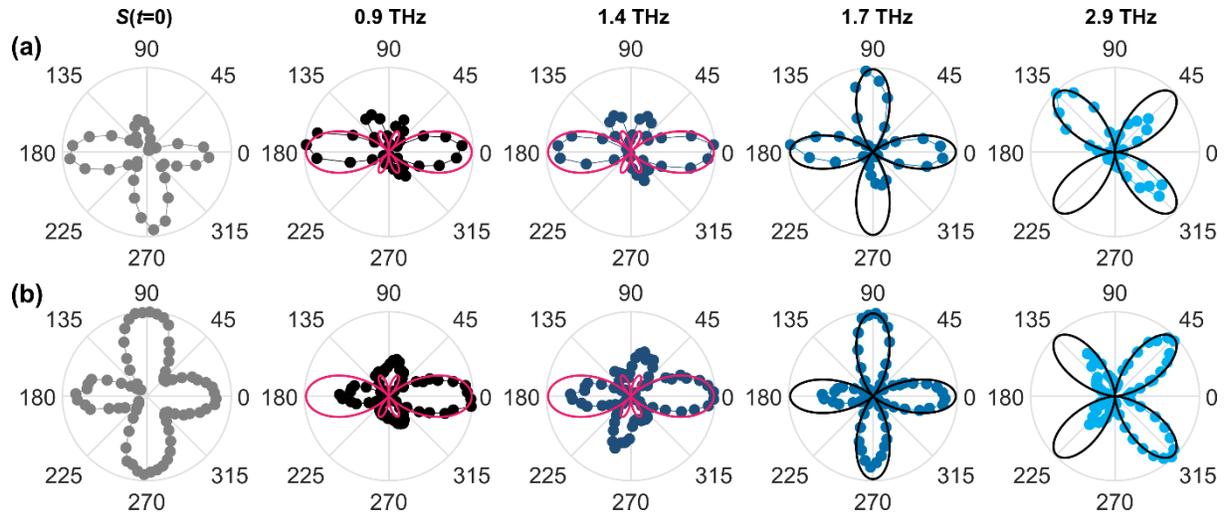

**Figure S7**. Instantaneous response amplitude $S(t=0)$, and mode amplitudes extracted from the Fourier transforms of the transient birefringence traces as a function of the sample azimuthal angle for two different $n=1$ flakes in **a)** and **b)**. In both cases, $\psi = 0°$ corresponds to long crystal axis parallel to the THz polarization. The magenta and black lines show the theoretically calculated coherent phonon signal amplitude assuming IR driving and Raman probing, or Raman driving and Raman probing, respectively.

**Section S8. Balancing artifacts in azimuthal angle scans**

Due to our experimental configuration, full azimuthal angle scans could not be performed while adjusting the balancing optics (quarter- and half-wave plate) in the detection setup for every angle $\psi$. Balancing was therefore done for $\psi = 0°$ and the positions of the balancing optics were left unchanged during the scan. Intrinsic sample birefringence in the 800 nm range may lead to slight changes in balancing conditions with azimuthal angle, which in turn could influence the signal amplitude. We estimate the magnitude of this effect comparing a scan where the configuration of the balancing optics was left unchanged from balancing at $\psi = 0°$ and a measurement where for each $\psi$, the balancing was adjusted manually. **Figure S8a** shows the non-normalized signals acquired in both configurations. **Figure S8b** compares the signal Fourier transforms for signals divided by the sum of the voltage on the two diode channels, which was acquired for every angle $\psi$ for each channel, and normalized so that the instantaneous signal amplitude is equal for the balanced and the unbalanced scan at $\psi = 0°$. Figure S8 (c) compares the azimuthal angle dependence of the instantaneous peak amplitude (normalized to be equal for both scans at $\psi = 0°$) and the FT spectral amplitude at the peaks at around 0.9 THz, 1.7 THz and 2.9 THz. The differences are negligible and do not influence our mode symmetry analysis and conclusions.

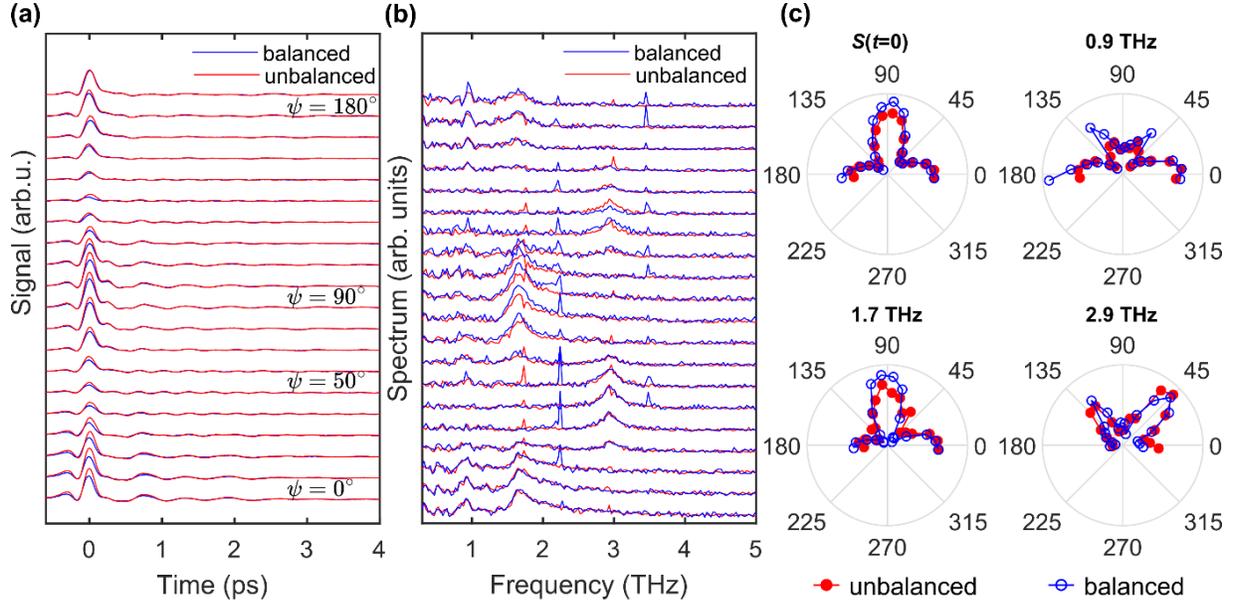

**Figure S8**. **a)** Transient birefringence signals for different sample azimuthal angles at 10° intervals measured with manual rebalancing at every angle (blue) and without (red). **b)** Corresponding FTs of the oscillatory part of the traces, normalized to match the instantaneous response magnitude at $\psi = 0°$ between the traces. **c)** Magnitude of the instantaneous response peak and the FT spectral amplitude at the 0.9 THz, 1.7 THz and 2.9 THz dominant modes as a function of azimuthal angle extracted from the measurements with and without rebalancing.

### Section S9. Phonon lifetime estimation

We estimate the phonon lifetimes based on the linewidths extracted from frequency-domain fits of Lorentzian functions to the respective peaks in the squared Fourier transform amplitude spectrum. As a cross-check we compared time- and frequency-domain fits. **Figure S9a** shows the fitting of the oscillatory signal in time domain with an exponentially damped sine function:

$$S(t) = A \cdot \exp(-\zeta_A t) \cdot \sin(2\pi f_0 t + \varphi) + B \tag{S7}$$

for an exemplary trace acquired on the *n*=1 sample at 150 K for the azimuthal angle $\psi = 45°$, after substracting an exponential tail related to the instantaneous response and propagation effects. We assume here that the amplitude decay time $\tau_A = \zeta^{-1}$ corresponds to twice the phonon lifetime τ (assuming the absence of elastic scattering leading to pure dephasing), $\tau_A$ =2τ. From the time domain fit, we extract the frequency and lifetime of the single mode as $f_0 = 3.062 \pm 0.004$ THz and $\tau = 1.9 \pm 0.2$ ps. Figure S9 (b) shows fitting of the squared Fourier transform spectral amplitude of the same trace with a single Lorentzian function:

$$S(f) = C \cdot \frac{\frac{\Gamma}{2}}{(f - f_0)^2 + (\Gamma/2)^2} + D, \tag{S8}$$

where $\Gamma$ corresponds to the peaks' full width at half maximum. Phonon lifetime is given by $\tau = 1/(2\pi\Gamma)$. From this fit, we obtain the mode frequency $f_0 = 3.066 \pm 0.003$ THz and a phonon lifetime of $\tau = 2.0 \pm 0.2$ ps. The results between time and frequency domain fitting agree very well, as expected when the signal amplitude fully decays within the measurement window in time domain. We therefore use only the frequency domain fitting for the remaining analysis.

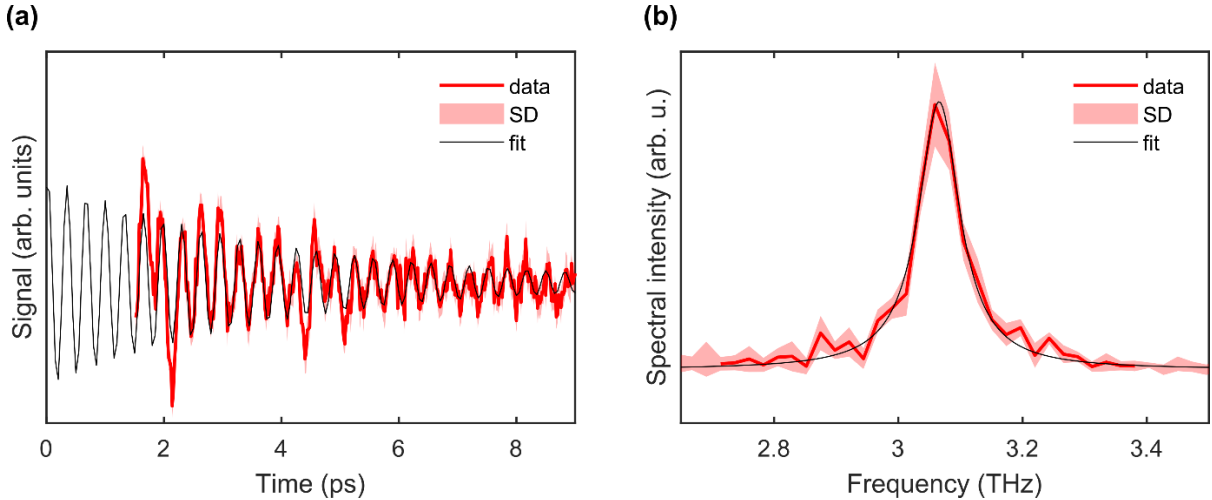

**Figure S9. a)** Oscillatory part of transient birefringence signal for *n*=1 sample at 150 K for the azimuthal angle $\psi = 45°$ (red) and damped sine function fit (black). The red shaded area corresponds to the standard deviation of the experimental data. **b)** Corresponding intensity spectrum (squared FT spectral amplitude) and Lorentzian fit in the range of interest.

Figure S10 shows exemplary frequency-domain fits of the low-frequency modes for *n*=1 at $\psi = 0°$ at (a) 80 K and (b) 150 K. We use a sum of three Lorentzian functions to fit the spetra in the 0.5-2 THz range. At 80 K, the lowest frequency peak shows a slight splitting, which we do not take into account for the purpose of our analysis.

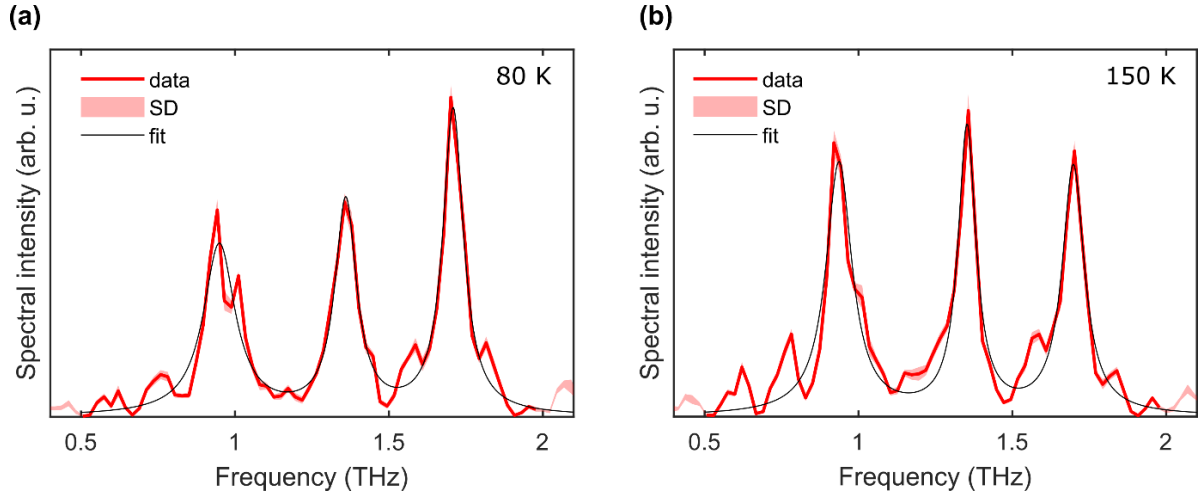

**Figure S10**. Intensity spectra for the *n*=1 sample at $\psi = 0°$ and fits of the three low-frequency phonon modes at **a)** 80 K and **b)** 150 K.

Figure S11 (b) illustrates the relative differences in mode lifetimes by comparing FT amplitude spectra of two room temperature transient birefringence traces for the *n*=1 sample at $\psi = 0°$ and $\psi = 45°$ (panel **a**), taken after cutting off the data at different pump-probe delay times. For $t > 0.5$ ps, the low-frequency mode lineshape for $\psi = 0°$ is distorted and resembles a Fano profile, due to the interference of the resonance and nonresonant terms in the nonlinear signal[8] (originating from the ionic and instantaneous electronic response, respectively). For $t > 1.5$ ps, the interference effects disappear and both low-frequency modes and the 2.9 THz mode are still detectable in the signal, while at longer times $t > 5$ ps only the low-frequency cage modes contribute to the oscillatory response.

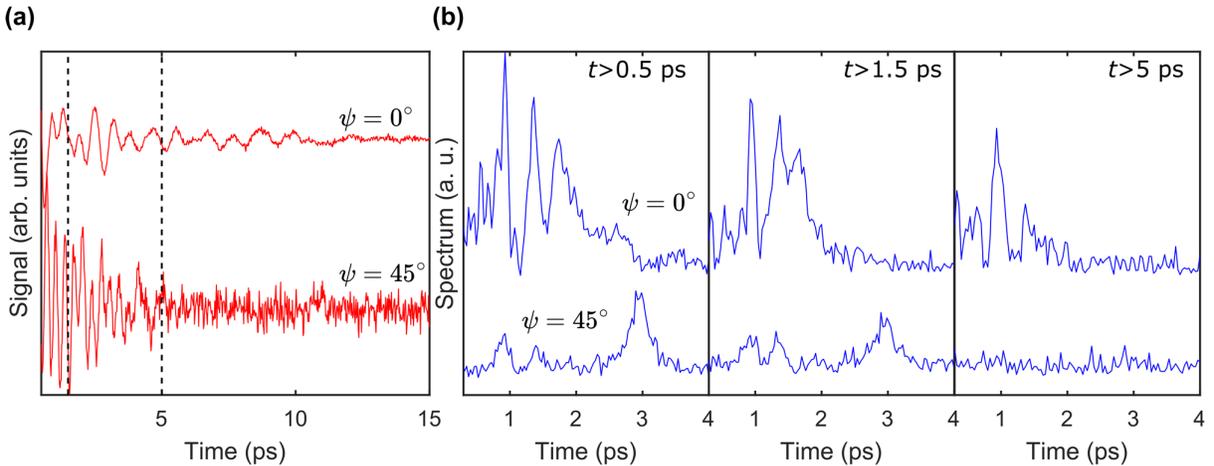

**Figure S11**. **a)** Transient birefringence signal of the *n*=1 sample at different azimuthal angles. Dashed lines mark the cutoff times. **b)** Corresponding FT spectral amplitude of the traces cut off at different times for the two azimuthal angles.

**Section S10. Temperature dependence of phonon lifetime and frequency**

Phonon lifetimes and frequencies are obtained based on fitting a Lorentzian function to the norm-squared of the FT amplitude in frequency domain. We assume that the observed decay of the coherent phonon population is mainly governed by inelastic defect-phonon and phonon-phonon scattering, and not by pure dephasing mechanisms, such as electron-phonon (elastic) scattering.[9] In this case, the inverse of the FWHM of the peak for a particular mode gives a good approximation of its lifetime.[9–11] We compare the observed shift and broadening of the phonon modes as a function of temperature $T$ with predictions of a model which considers only the simplest case of a symmetric 3-phonon anharmonic decay.[12,13] The shift of a phonon mode with temperature in this approximation is the sum of the term related to lattice expansion $\Delta f_0(T)$ as well as the one due to the phonon-phonon scattering $\Delta f_{\mathrm{ph},3}(T)$:[12]

$$\Delta f(T) = \Delta f_0(T) + \Delta f_{\mathrm{ph},3}(T), \tag{S9}$$

$$\Delta f_{\mathrm{ph},3}(T) = A \left( 1 + \frac{2}{e^{\frac{hf_0}{2k_{\mathrm{B}}T}} - 1} \right). \tag{S10}$$

$A$ is the anharmonic constant, $k_{\mathrm{B}}$ the Boltzmann constant, $h$ the Planck constant and $f_0$ the phonon frequency in the harmonic approximation at 0 K. We neglect the influence of lattice expansion on the mode frequencies as we do not have detailed information on the evolution of the lattice constants, and also given that the modes have largely in-plane character and the in-plane lattice expansion as a function of temperature in 2D layered perovskites is not significant and strikingly weaker than in the out-of-plane direction.[14] We therefore fit the observed shift using only the second term in expression (1) and obtain the following parameters (**Table S1**):

| Mode | 0.9 THz | 1.4 THz | 1.7 THz | 3 THz |
|---|---|---|---|---|
| $A$ [THz] | -0.00010 | -0.00014 | -0.00029 | -0.0036 |
| $f_0$ [THz] | 0.94 | 1.37 | 1.72 | 3.15 |

**Table S1**. Parameters obtained from the phonon frequency shift fitting.

The same 3-phonon scattering model predicts the evolution of the phonon linewidth (and corresponding lifetime) as a function of temperature:

$$\Gamma_3(T) = \Gamma_0 + B \left( 1 + \frac{2}{e^{\frac{hf}{2k_{\mathrm{B}}T}} - 1} \right) \tag{S11}$$

For the $\Gamma(T)$ fitting, we can assume the frequency $f = f_0$ as a constant parameter equal to the un-renormalized frequency $f_0$ at 0 K, as the change of $f$ with temperature is relatively small. $\Gamma_0$ is the temperature-independent broadening due to impurity scattering,[15] and the anharmonic constant $B$ corresponds to the intrinsic linewidth at $T=0$. We performed this analysis only for the mode at around 3 THz, as the other modes do not present a clear enough dependence. The fitting parameters obtained are given in **Table S2**:

| Mode | 3 THz |
|---|---|
| $B$ [THz] | 0.0023 |
| $f_0$ [THz] | 3.16 |
| $\Gamma_0$ [THz] | 0.017 |

**Table S2**. The parameters obtained from the phonon linewidth fitting.

While in reality, contributions from higher order and asymmetric scattering processes,[16,17] as well as corrections due to lattice expansion may be non-negligible, the simple model qualitatively captures the observed mode behavior. We do not assign the phonon branches into which the original optical modes scatter, which could in principle either be both acoustic (for example the symmetric Klemens mechanism[13]) and optical, as we lack information on the full phonon dispersion of the complex RPP material.

**Section S11. Raman tensor assignment and probe polarization dependence**

We base our Raman tensor element assignment for the 1.4, 1.7 and 2.9 THz modes on fitting the polarized spontaneous Raman scattering data (see **Section S14**). Using the same tensors to describe the probing process and their real parts to describe the driving process allows us to well reproduce the transient birefringence experimental data. **Table S3** summarizes the Raman tensor assignment and the driving mechanisms of the different modes. Note that as given in Table S3 describe only the general form of the tensors, and to reproduce the quantitative results a scaling factor multiplying the tensor should be used to account for different intensities of the modes. For the 0.9 THz mode we assume the same form of the Raman tensor as for the 1.7 THz mode, as its low intensity in spontaneous Raman scattering hinders precise fitting of the tensor elements.

To confirm the assignment of the mode symmetries, we measured the transient birefringence signal for the $n$=1 sample as function of the probe polarization direction (changed by rotating a half-wave plate in the incoming probe beam) for three different, fixed flake orientations, as shown in **Figure S12**. To calculate the signal, we assigned Raman symmetries and driving mechanisms as summarized in Table S3. Comparing the experimental signal of the four dominant vibrational modes (taking into account both amplitude and sign) with theoretical calculations we obtain very good agreement, as can be seen in **Figure S12 c, f, i.** The very weak signal of the 0.9 THz mode, observed for the flake oriented perpendicular to the THz polarization (Figure S12f), may be present due to imperfect polarization and flake orientation, or represent a weak contribution from the Raman-type driving, not taken into account in our calculation.

| Mode | 0.9 THz | 1.4 THz | 1.7 THz | 2.9 THz |
|---|---|---|---|---|
| Driving | IR | IR | Raman | Raman |
| Raman tensor | $\mathbf{R} = a \begin{bmatrix} 1 & \frac{4}{3}e^{i\pi/2} \\ \frac{4}{3}e^{i\pi/2} & -1 \end{bmatrix}$ | $\mathbf{R} = a \begin{bmatrix} 1 & e^{i\pi/2} \\ e^{i\pi/2} & -1 \end{bmatrix}$ | $\mathbf{R} = a \begin{bmatrix} 1 & \frac{4}{3}e^{i\pi/2} \\ \frac{4}{3}e^{i\pi/2} & -1 \end{bmatrix}$ | $\mathbf{R} = b \begin{bmatrix} e^{i\pi/2} & -\frac{8}{9} \\ -\frac{8}{9} & -e^{i\pi/2} \end{bmatrix}$ |

**Table S3**. Driving mechanisms and Raman tensors for the different vibrational modes.

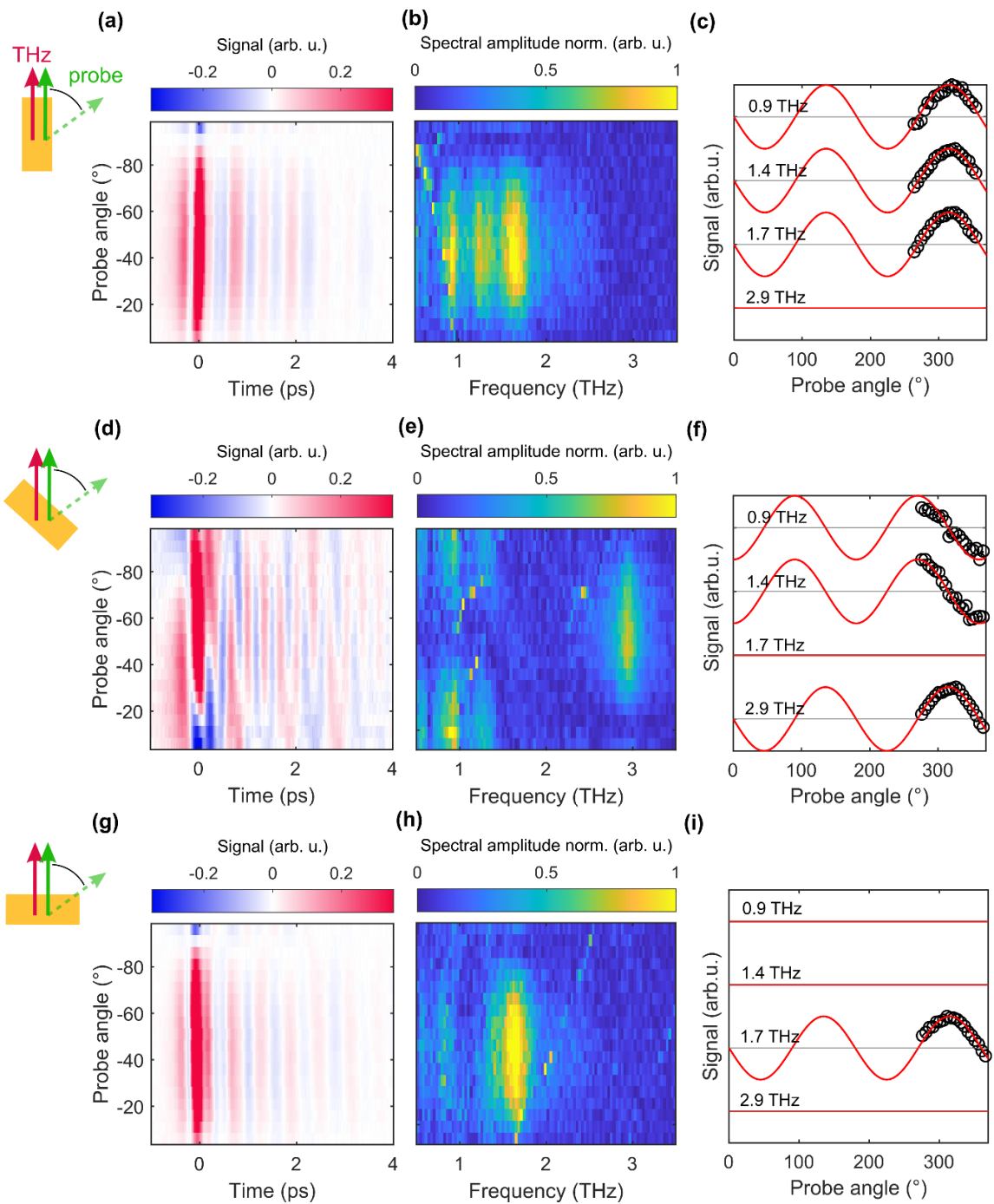

**Figure S12**. Probe polarization scans for flake long axis oriented parallel to the THz (**a)-c)**), at −45° (**d)-f)**) and perpendicular (**g)-i)**). Panels **a)**, **d)**, **g)**: transient birefringence signal as a function of probe polarization angle. Panels **b)**, **e)**, **h)**: signal Fourier transform amplitude. Panels **c)**, **f)**, **i)**: experimental (black dots) and theoretically calculated (red lines) signal amplitude as a function of probe polarization angle for the different Raman-active modes.

**Section S12. Raman and IR driving mechanism simulations for the 0.9 and 1.4 THz modes**

To confirm the assignment of the symmetries and driving mechanisms for the 0.9 THz and 1.4 THz mode, which show the complex multi-lobed transient birefringence pattern as a function of the sample azimuthal rotation angle, we calculated the driving force $F_{\mathrm{dr}}$, probing sensitivity $S_{\mathrm{pr}}$, and transient birefringence signal using alternative forms of the mode's Raman tensors. We compared the results obtained using the tensors:

$$\mathbf{R}^\alpha = a \begin{bmatrix} 1 & \frac{4}{3}e^{i\pi/2} \\ \frac{4}{3}e^{i\pi/2} & -1 \end{bmatrix}, \qquad \mathbf{R}^\gamma = \begin{bmatrix} a & 0 \\ 0 & c \end{bmatrix}, \tag{S12}$$

with real $c \neq a$, $c>0$, $a>0$, and assumed either Raman, or IR driving as described in detail in Section S18. The comparison between the theoretical signals calculated using $\mathbf{R}^\alpha$ and $\mathbf{R}^\gamma$ shows that also this alternative Raman tensor including *xy* in-plane anisotropy cannot explain the observed angle dependence in a Raman driving – Raman probing picture. Figure S13 shows that only the combination of IR driving and probing via $\mathbf{R}^\alpha$ reproduces the experimental observations, in particular the signal sign changing as a function of $\psi$.

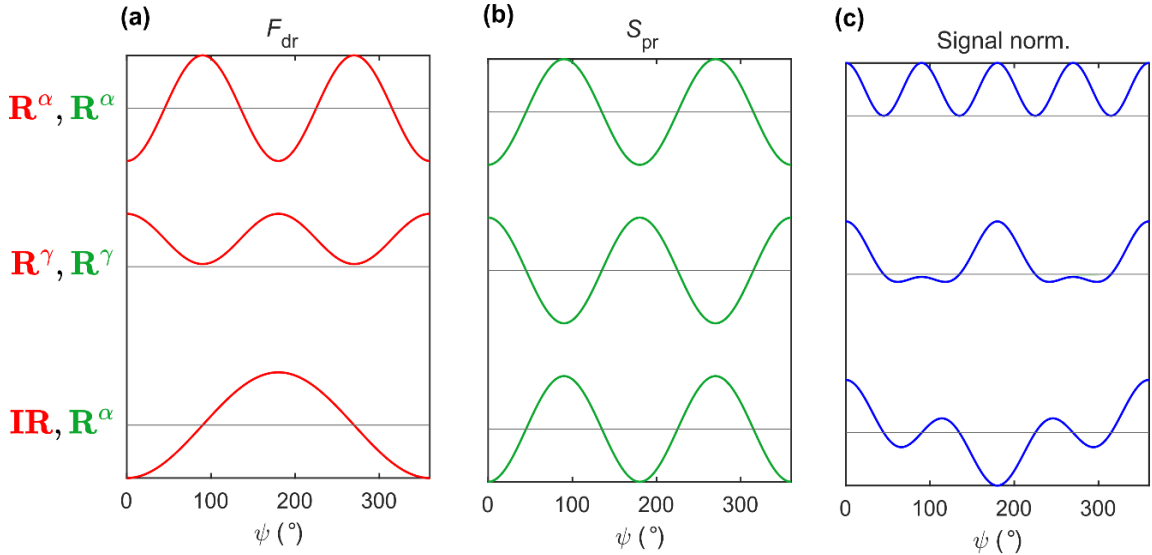

**Figure S13**. Calculated **a)** driving force, **b)** probing sensitivity, **c)** transient birefringence signal as a function of sample azimuthal angle for three different combinations of driving and probing mechanisms: Raman-type driving and probing described by $\mathbf{R}^\alpha$ (top)**,** Raman-type driving and probing described by $\mathbf{R}^\gamma$ (middle) and IR driving and Raman-type probing described by $\mathbf{R}^\alpha$ (bottom). The black lines indicate zero of the calculated functions. The signal corresponds to the product of $F_{\mathrm{dr}}$ and $S_{\mathrm{pr}}$. We use an unphysically high anisotropy ratio of the $\mathbf{R}^\gamma$ tensor elements *c*=20*a* to demonstrate the effect more clearly.

**Section S13. Azimuthal angle scans for *n*=2,3**

We measured transient birefringence signals for samples with *n*=2,3 inorganic layers as a function of the crystal azimuthal angle. Angle scans for *n*=1,2 and 3 are compared in **Figure S14 a-f**. For all the samples, the signal originating from the lowest-frequency modes shows a dependence on the field polarity, suggesting IR-type driving. This can be visualized by the imaginary part of the Fourier transform **d-f**, as well as by comparing the inverse Fourier transforms after spectral filtering upon a 180° rotation, shown for the dominant low frequency modes in **Figure S14 j-l**. The balancing was not adjusted during the scans. While, as shown in section S8, the effect of this is mostly negligible for *n*=1, we did not perform such cross-checks for *n*=2,3, therefore the exact relative intensities measured in the angle scan may carry some error. The relative intensities at different angles $\psi$ also carry an error due to sample inhomogeneities and imperfect centering of the rotation axis in the experimental setup.

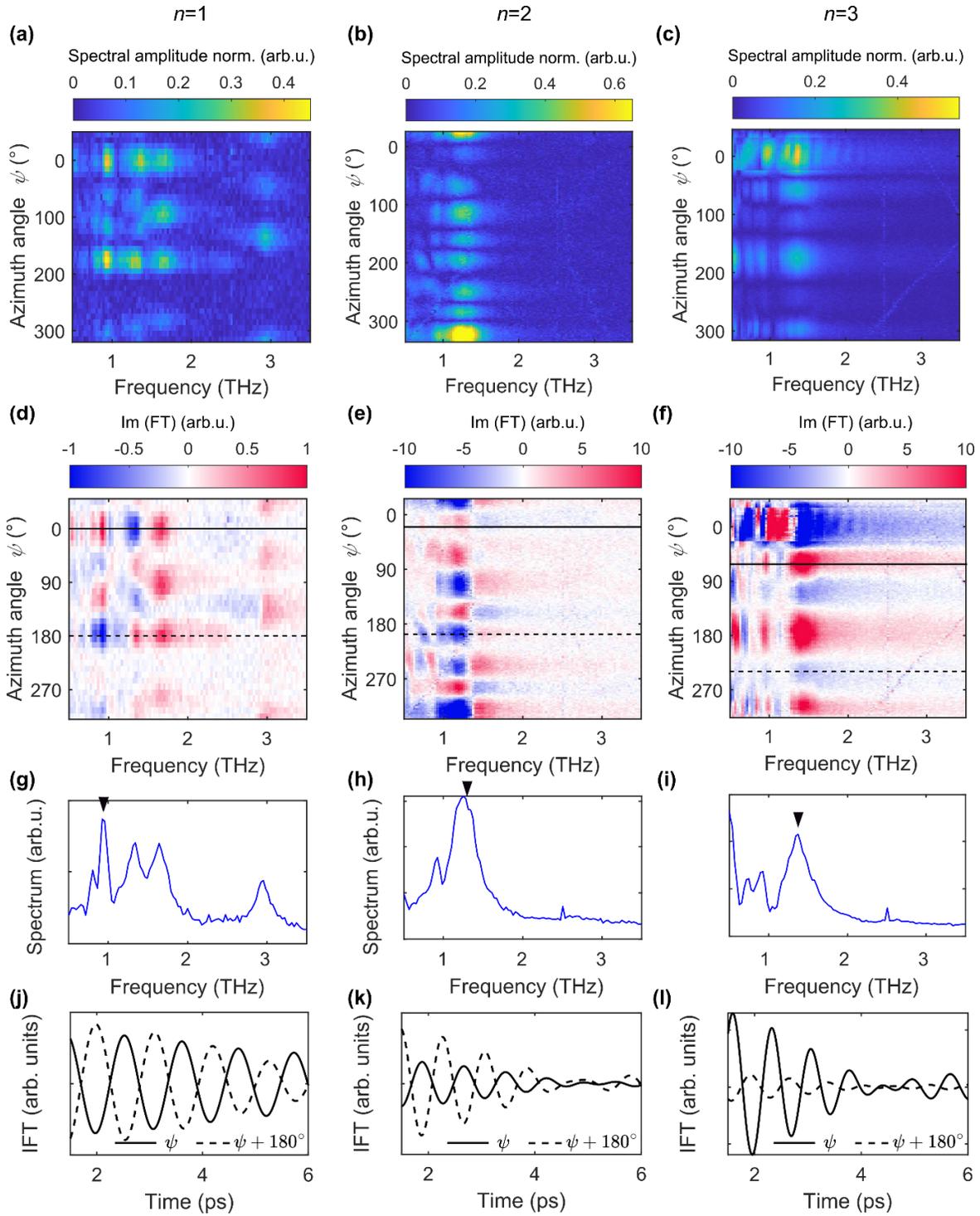

**Figure S14. a)-c)** spectral amplitude of the transient birefringence signal FT for *n*=1,2,3 samples as a function of azimuthal angle and **d)-f)** the corresponding imaginary part of the FTs. **g)-i)** Spectra averaged over 180°. **j)-l)** Inverse Fourier transforms calculated after spectral filtering around the dominating peaks marked with black triangles in panels g)-i), for two angles marked in panels d)-f) with the solid and dashed lines.

## Section S14. Raman spectroscopy

We characterized the *n*=1 and *n*=2 using polarization-resolved spontaneous Raman scattering spectroscopy. The measurements were performed under ambient conditions, using a 647 nm excitation laser. **Figure S15** shows static Raman spectra measured on two different spots for each of the samples: with unpolarized detection for arbitrary excitation laser polarization (grey, the sum of the co- and cross-polarized spectra averaged over a full polarization scan (black), as well as the azimuthal angle-averaged transient birefringence spectrum squared (blue). The Raman spectra show good agreement between the different spots.

For *n*=1, the 0.9 THz mode, prominent in the transient birefringence spectrum, is negligible in the spontaneous Raman spectra. This can be due to its weak Raman activity, its dominance in the transient birefringence experiments being related to the very efficient linear direct driving. Relative differences in mode amplitude between the two experiments may be related to the driving force spectral profile, as discussed in Section S4. Certain modes appear absent in the transient birefringence spectrum (2.15 THz and 3.4 THz), which may be due to the balanced detection scheme being uniquely sensitive only to anisotropic modes.

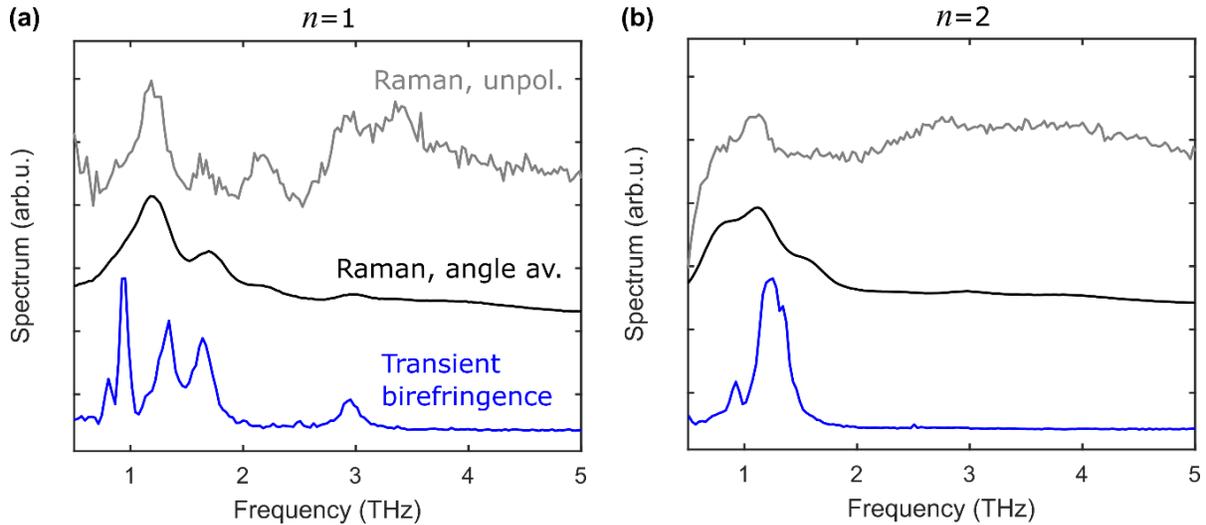

**Figure S15**. Comparison between spontaneous Raman scattering and transient birefringence measurements for **a)** *n*=1 and **b)** *n*=2. Grey: Raman spectrum for arbitrary excitation polarization and unpolarized detection. Black: sum of the Raman signal collected in co- and cross-polarized configuration, averaged over the sample azimuthal angles $\psi$: $\Sigma_\psi(I_\parallel(\psi)) + \Sigma_\psi(I_\perp(\psi))$. Blue: squared transient birefringence spectrum, averaged over $\psi$: $\Sigma_\psi |\mathrm{FT}(S(\psi))|^2$.

**Figure S16** shows intensity maps of the Raman spectra measured in the co- and cross-polarization configuration as a function of the laser in-plane polarization angle for the two samples. The marked dominant modes show 2-fold and 4-fold symmetry, with maxima for laser polarization along the long flake edge direction, or at $45°$ to it.

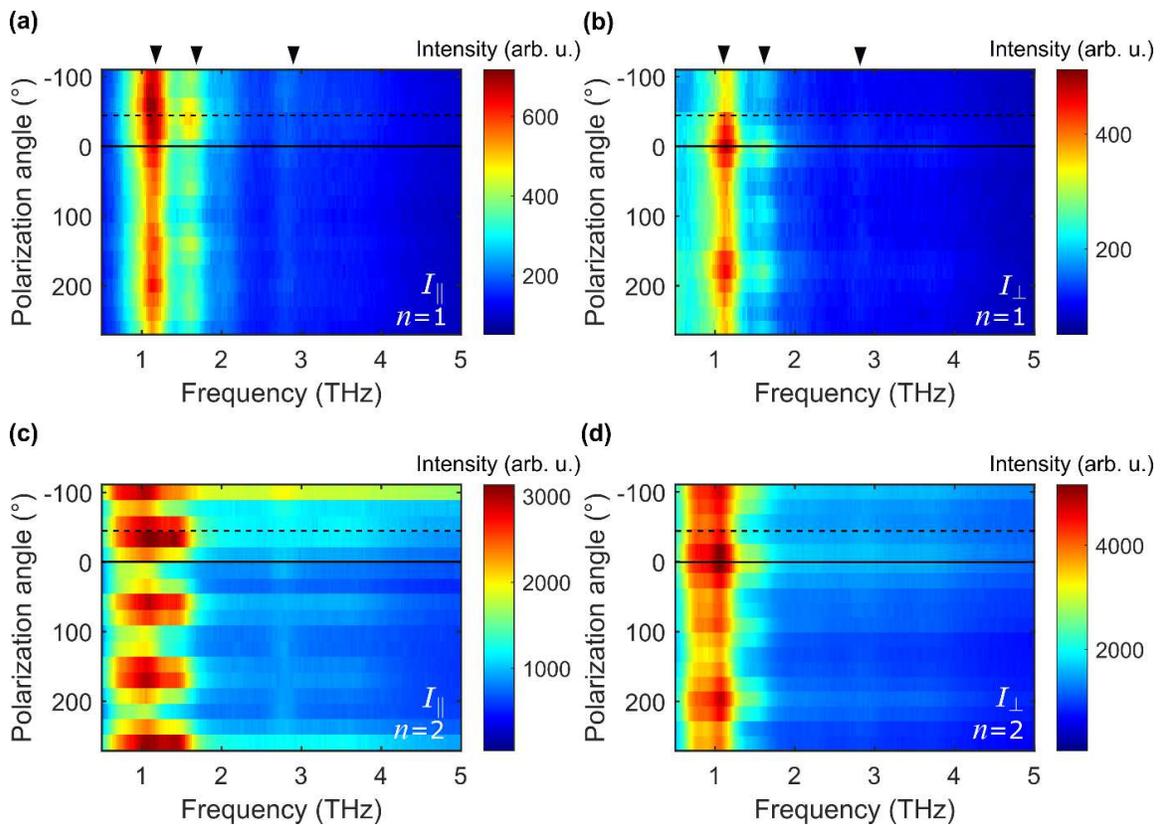

**Figure S16**. Polarized Raman spectra for *n*=1 **a)** and **b)**) and *n*=2 **c)** and **d)**) as a function of the excitation polarization angle. Maps in a) and c) show to co-polarized spectra, and b) and d) cross-polarized spectra. Dominant modes at ~1.2 THz, 1.7 THz and 2.9 THz are marked for *n*=1. The solid and dashed lines in all panels correspond to excitation polarization at $0°$ and $45°$ to the long crystal axis, respectively.

**Figure S17** shows Raman spectra detected in both polarization configurations with the laser polarized at $0°$ and $45°$ to the long flake edge, corresponding to the lines in Figure S16, and squared transient birefringence spectra for the THz at $0°$ and $45°$ to the long flake edge.

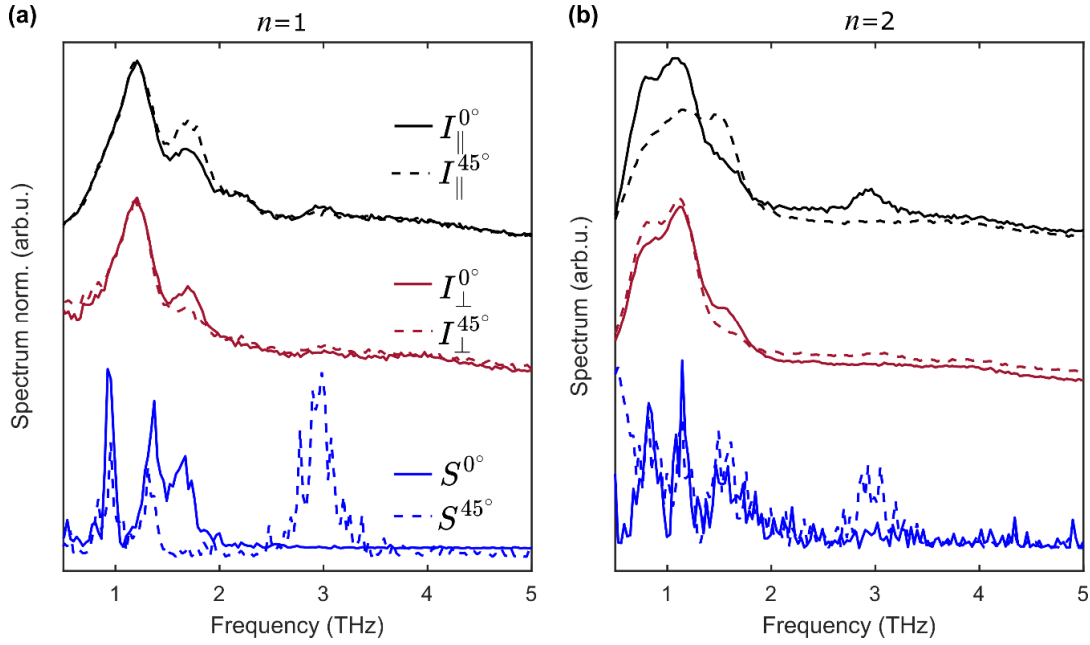

**Figure S17.** Co- (black) and cross-polarized (red) Raman spectra and squared transient birefringence spectra (blue) for $n=1$ **a)** and **c)**) and $n=2$ **b)** and **d)**). Solid lines show Raman spectra acquired at excitation polarization at $0°$ to the flake long edge, and transient birefringence measured for THz polarization parallel to the flake long axis. Dashed lines show spectra measured at the respective angles of $45°$ to the flake long edge.

**Figure S18 a, b, f, g** shows polar plots of the dominant 1.7 THz and 2.9 THz Raman peak experimental intensities as a function of the excitation polarization angles in the co- and cross-polarized configuration compared with theoretical calculations based on assigning the tensors of the modes as $\mathbf{R}^\alpha$ and $\mathbf{R}^\beta$, as given in the Main text and Table S3, respectively. The Raman peak intensity measured in those configurations is given by $I_{co} \propto |\hat{e}_i \mathbf{R} \hat{e}_i|^2$ and $I_{cross} \propto |\hat{e}_i^\perp \mathbf{R} \hat{e}_i|^2$, where $\hat{e}_i$ is a vector along the polarization direction of the incident laser beam and $\hat{e}_i^\perp$ perpendicular to it and $\mathbf{R}$ the mode Raman tensor. The tensor elements are obtained based on fitting the Raman intensity angle dependence up to a phase factor of $e^{i\pi/2}$. The final phase is assigned to obtain agreement with the transient birefringence experimental results.

**Figure S18 c, d, i, h** shows the azimuthal angle dependence of the driving force amplitude and probing sensitivity, as derived in **Section 17**. The probing sensitivity is calculated using $\mathbf{R}^\alpha$ and $\mathbf{R}^\beta$ for the two modes. We use the real part of the respective tensors to describe the driving process. **Figure S18 e, j** shows the corresponding experimentally measured and calculated TKE signal amplitude, as described in **Section S18**.

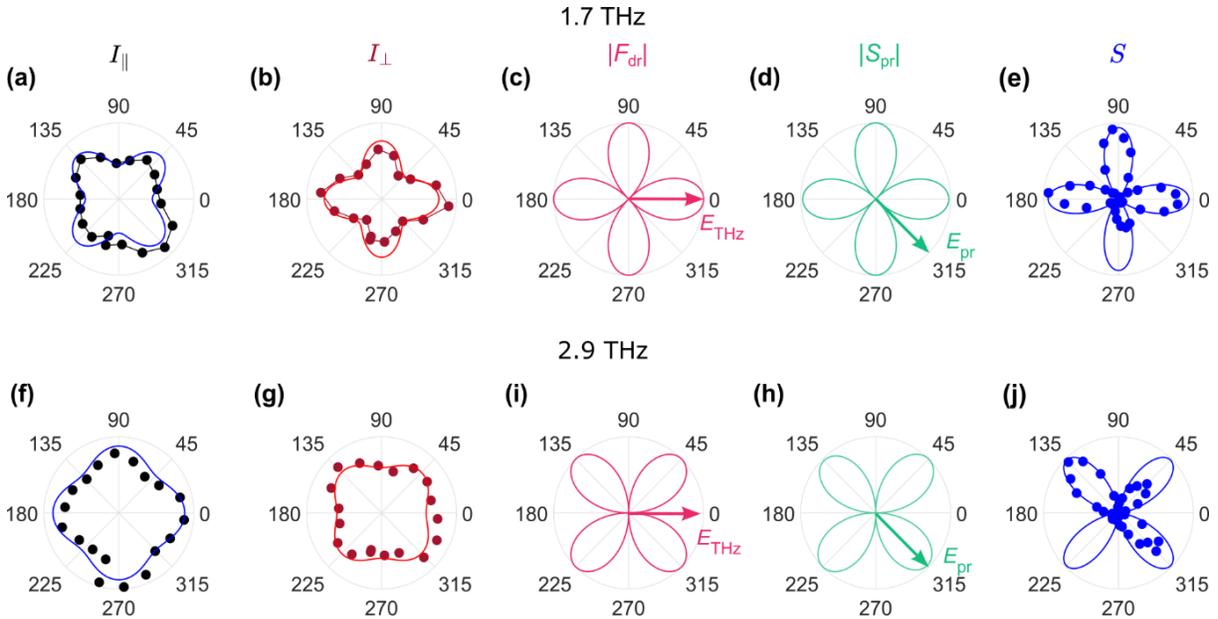

**Figure S18. a)-e)** 1.7 THz mode: Co-polarized ($I_\parallel$, black, **a)**) and cross-polarized ($I_\perp$, red, **b)**) Raman intensity, driving force amplitude **c)**, probing sensitivity **d)** and TKE signal amplitude (blue, **e)**). Points show experimental data and solid lines calculations based on the assigned Raman tensors. $\psi = 0°$ corresponds to long flake edge and the arrows mark the THz pump and probe polarization directions. **f)-j)** 2.9 THz mode.

### Section S15. Transient birefringence measurements on PEA halide salt crystals

**Figure S19a** shows the transient birefringence traces acquired at room temperature on PEAI and PEABr crystals, compared with the trace for the $n$=1 perovskite sample. For the ligand-halide crystals, thin samples (on BK7 glass substrates) and thicker crystals (**Figure S19d**) are compared. While clear oscillations are visible in both PEABr traces on the ps timescale, PEAI does not show distinguishable coherent phonon signatures. This may be related to much higher structural disorder, as can be seen in the optical images in **d**, as PEAI in contrast to PEABr does not form well oriented crystals on the substrate but rather disordered thin films. PEAI thick crystals also show an incoherent response after THz excitation, which does not fully decay in our measurement window of tens of ps, likely due to heating effects and slow heat dissipation in PEAI. The instantaneous response in the thin films shows an unipolar character, following $E_{THz}^2$, in PEABr and $n$=1 perovskite and a bipolar form in PEAI, reminiscent of earlier measurements in butylammonium ligands.[18] In thick crystals, the response may be strongly affected by anisotropic propagation effects[2] and is therefore not representative. The differences between THz-driven dynamics between PEAI, PEABr and perovskite crystals point to a strongly synergistic character of the response. Panel **b** shows the spectra of the oscillatory response in PEABr and (PEA)$_2$PbI$_4$ thin films. The clearly distinct peaks observed in both materials demonstrate, that the low-frequency rigid-body modes of the organic ligand molecules are highly sensitive to the local environment and the entire hybrid crystal structure. The vibrational response is strongly anisotropic for the highly ordered PEABr crystals, as shown in the azimuthal angle scan in panel **c**, which evidences the tendency of the PEA molecules to form highly ordered structures.

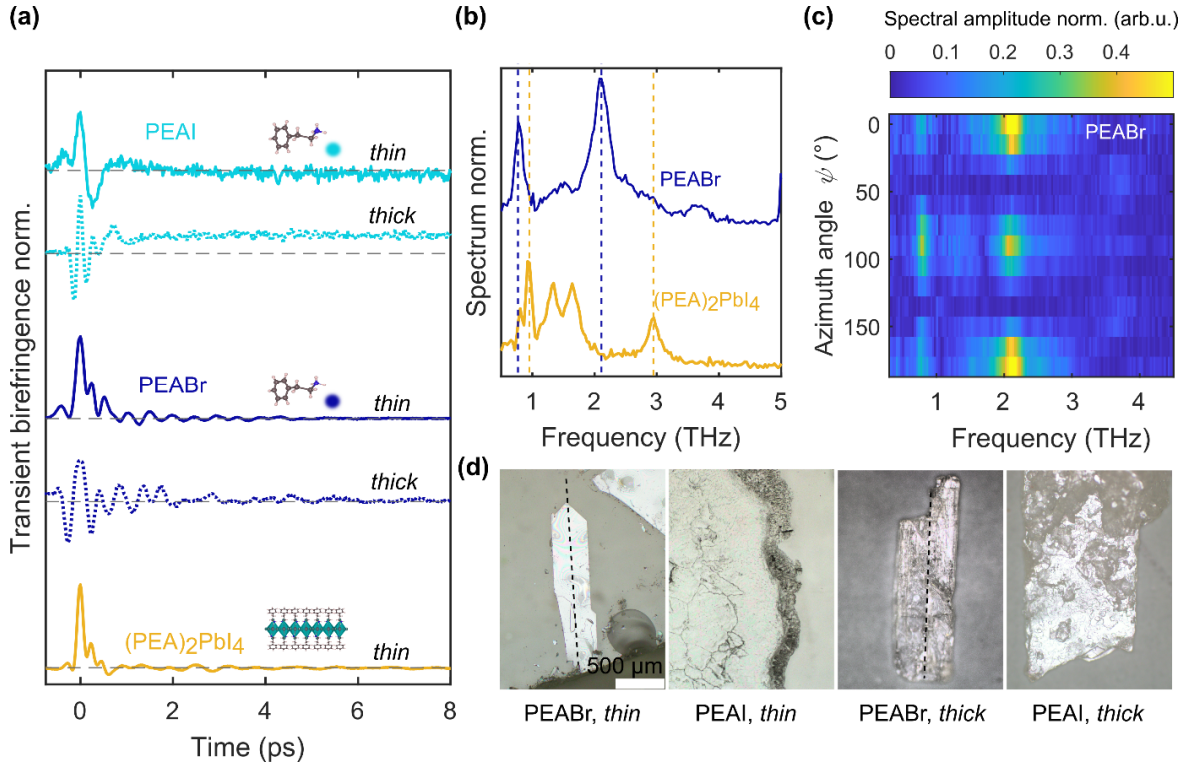

**Figure S19**. **a)** Normalized transient birefringence traces for thin and thick PEAI and PEABr crystals and the $n$=1 perovskite thin crystal sample. **b)** Normalized, azimuthal angle-averaged transient birefringence spectra for PEABr and $n$=1 thin crystals. **c)** Map of spectral amplitude for PEABr as a function of the sample azimuthal angle. **d)** Optical images of the measured ligand-halide crystals.

### Section S16. THz transmission

We measured THz transmission through a thick, free-standing $n$=1 perovskite crystal for different sample azimuthal orientations via electro-optic sampling using a 100 μm ZnTe detection crystal. **Figure S20a** shows the Fourier transforms of the as-measured traces after clipping the signal in time domain to filter out pump and probe reflections. **Figure S20b** shows the relative transmittance at the respective angles, calculated by taking the ratios of the squared sample and reference signal spectral amplitudes, compared with an angle-averaged $n$=1 thin crystal transient birefringence spectrum. The relative transmittance is only meaningful for frequencies < 1.5 THz because of the signal to noise ratio at higher frequencies. A pronounced dip in the relative transmittance spectrum at around 0.9 THz corresponds to the dominant THz-induced transient birefringence peak, which we assigned to a simultaneously IR- and Raman-active mode.

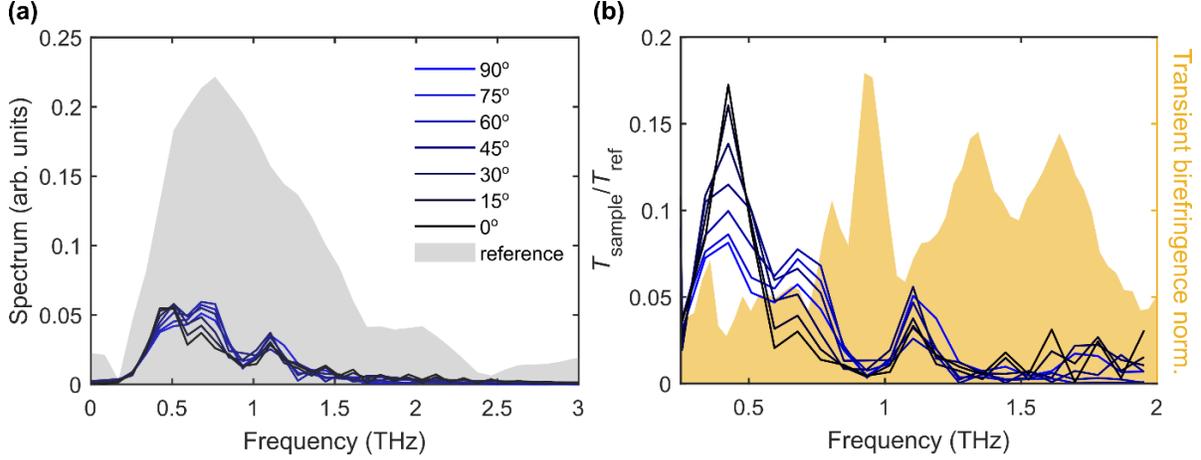

**Figure S20. a)** Spectral amplitude of electro-optic sampling signal, acquired without the sample (reference, grey area) and for the *n*=1 sample at different azimuthal angle orientations, where $\psi = 0°$ corresponds to the long crystal axis parallel to the incident THz polarization. **b)** Relative transmittance $T_{\text{sample}}/T_{\text{ref}}$, compared with angle-averaged transient birefringence spectrum.

**Section S17. Theoretical description of coherent phonon driving and probing**

*S17.1. Coherent phonon driving*

In the discussion of our experimental results, we consider two types of coherent phonon generation mechanisms: linear and nonlinear photonic Raman-type driving.

Linear driving is possible for IR-active (polar) modes with resonance frequencies within the driving field spectrum. IR-active modes are associated with a change of the macroscopic polarization per unit cell with the mode normal coordinate, which means that the mode has nonzero effective charge. The *Born effective charge tensor* $Z^*_{n,ij}$ contribution of the *n*-th ion is defined as:[19]

$$Z^*_{n,ij} = V_0 \left.\frac{\partial \mathcal{P}_j}{\partial u_{n,i}}\right|_{E=0}, \tag{S13}$$

where $V_0$ is the unit cell volume, $u_{n,i}$ the displacement of ion *n* along the spatial direction *i* and $\mathcal{P}_j$ the macroscopic polarization in the unit cell along the direction *j* for zero macroscopic electric field $E$. Modelling coherent phonons as driven, damped harmonic oscillators and assuming a harmonic lattice potential, in frequency domain we write the equation of motion for an IR-active mode with the normal mode coordinate $Q_{\text{IR}}(\Omega)$ under external electric field $\boldsymbol{E}^{\text{dr},1}(\omega_1)$ of the driving pulse as:[20]

$$(\Omega_{\text{IR}}^2 - \Omega^2 - i\Gamma\Omega)Q_{\text{IR}} = F_{\text{dr}}^{\text{IR}}(t) = \boldsymbol{\mathcal{Z}}^* \cdot \boldsymbol{E}^{\text{dr},1}(\omega_1), \tag{S14}$$

where $\Gamma$ is the damping coefficient, $\Omega_{\text{IR}}$ the mode eigenfrequency and $\boldsymbol{\mathcal{Z}}^* = \sum_n Z^*_n u_n$ is the *mode effective charge vector*[20] coming from the summation over the contributions of all the ions in the unit cell. $F_{\text{dr}}^{\text{IR}}$ is the driving force for the IR-active mode. Phonon driving in this case occurs via a one-field interaction.

Modes for which $\mathcal{Z}^* = 0$ can be driven via a two-field, Raman-type interaction. For these modes, the linear susceptibility $\chi^{(1)}(Q_\text{R})$ changes as a function of the normal mode coordinate and can be expanded into a Taylor series for small perturbations around the equilibrium position $Q_\text{R} = 0$:

$$\chi^{(1)}(Q_\text{R}) = \chi_0^{(1)} + \left.\frac{\partial \chi^{(1)}}{\partial Q_\text{R}}\right|_{Q_\text{R}=0} Q_\text{R} + \cdots, \tag{S15}$$

where the derivative $\left.\frac{\partial \chi_{ij}^{(1)}}{\partial Q_\text{R}}\right|_{Q_\text{R}=0} \equiv \mathbf{R}_{ij}$ defines the *Raman susceptibility tensor*.[21] The Raman tensor element values are in general frequency dependent due to the dispersion of $\chi^{(1)}$ and specific to a particular mode at frequency $\Omega_\text{R}$: $\mathbf{R}^{\omega,\Omega_\text{R}} = \left.\frac{\partial \chi^{(1)}(\omega)}{\partial Q_\text{R}}\right|_{Q_\text{R}=0}$. In the presence of a driving electric field component $\boldsymbol{E}^{\text{dr},1}$ of the frequency $\omega_1$, the induced nonequilibrium polarization $\boldsymbol{P}(Q_\text{R}, \boldsymbol{E}^{\text{dr},1})$ in the material is then given by:[21]

$$\boldsymbol{P}(Q_\text{R}, \boldsymbol{E}^{\text{dr},1}) = \varepsilon_0 \chi^{(1)}(\omega_1, Q_\text{R})\boldsymbol{E}^{\text{dr},1} = \boldsymbol{P}_\text{el} + \varepsilon_0 Q_\text{R} \mathbf{R}^{\omega_1,\Omega_\text{R}} \boldsymbol{E}^{\text{dr},1}, \tag{S16}$$

with the first non-resonant term $\boldsymbol{P}_\text{el} = \varepsilon_0 \chi_0^{(1)} \boldsymbol{E}^{\text{dr},1}$, corresponding to instantaneous electronic polarization, and the second term, modulated as a function of the Raman-active mode displacement. Here, the Raman susceptibility tensor $\mathbf{R}^{\omega_1,\Omega_\text{R}}$ is defined for the driving electric field frequency $\omega_1$, and the Raman phonon natural frequency $\Omega_\text{R}$. The potential energy $W$ of the nonequilibrium polarization under a second driving electric field component, $\boldsymbol{E}^{\text{dr},2}$, is given by $W = -\boldsymbol{E}^{\text{dr},2} \cdot \boldsymbol{P}(Q_\text{R}, \boldsymbol{E}^{\text{dr},1}) = \boldsymbol{E}^{\text{dr},2} \cdot \boldsymbol{P}_\text{el} + \varepsilon_0 Q_R \boldsymbol{E}^{\text{dr},2} \cdot \mathbf{R}^{\omega_1,\Omega_\text{R}} \boldsymbol{E}^{\text{dr},1}$. The driving force for the collective dynamics of a Raman-active vibrational mode from this is as follows:

$$F_{\text{dr}}^{\text{R}} = -\frac{\partial W}{\partial Q_\text{R}} = \varepsilon_0 \boldsymbol{E}^{\text{dr},2} \mathbf{R}^{\omega_1,\Omega_\text{R}} \boldsymbol{E}^{\text{dr},1}. \tag{S17}$$

As can be seen from Equation (S17), two field interactions are necessary to drive a Raman-active mode. We assume that the Raman tensor $\mathbf{R}^{\omega_1,\Omega_\text{R}}$ describing the coherent driving process has purely real elements. Under this assumption, using the real part of the Raman tensors obtained from fitting the spontaneous Raman results in the visible range as $\mathbf{R}^{\omega_1,\Omega_\text{R}}$, we obtain very good agreement of our theoretical description with experimental results. In our case, $\boldsymbol{E}^{\text{dr},1}$ and $\boldsymbol{E}^{\text{dr},2}$ are two THz electric field components at frequencies $\omega_1$, $\omega_2$ from within the bandwidth of the driving pulse. Considering energy conservation for the Raman process, the eigenfrequency of the coherently-driven Raman mode $\Omega_\text{R}$ has to be equal to $\Omega_R = (\omega_1 + \omega_2)$ or $\Omega_R = (\omega_1 - \omega_2)$ for a sum-frequency (analogous to two photon absorption[22]) or difference-frequency excitation process, respectively.[20,22] The equation of motion for the Raman-active mode then becomes:[20]

$$(\Omega_R^2 - \Omega^2 - i\Gamma\Omega)Q_R = F_{dr}^R = \varepsilon_0 \boldsymbol{E}^{dr,2}\boldsymbol{R}^{\omega_1,\Omega_R}\boldsymbol{E}^{dr,1}. \tag{S18}$$

In principle, another excitation mechanism can lead to an observed quadratic dependence of the mode amplitude on the driving THz electric field, the so-called ionic driving mechanism. In this pathway, the THz electric field directly excites IR-active modes which subsequently anharmonically couple to drive a Raman-active mode.[23] Distinguishing between the Raman-type photonic and ionic mechanism is beyond the scope of this work and for simplicity we focus here on the photonic mechanism which, in contrast to the ionic mechanism, is universally possible.

*S17.2. Coherent phonon detection*

Coherent phonon probing in our detection scheme relies on the modulation of the linear susceptibility $\chi^{(1)}(\omega_{pr}, Q_R)$ at the frequency $\omega_{pr}$ of the probing field by the collective Raman-active phonon motion. We again Taylor-expand the susceptibility as a function of the vibrational coordinate $Q_R$:

$$\chi^{(1)}(\omega_{pr}, Q_R) = \chi_0^{(1)}(\omega_{pr}) + \frac{\partial \chi^{(1)}(\omega_{pr})}{\partial Q_R} Q_R + \cdots. \tag{S19}$$

$\boldsymbol{R}^{\omega_{pr},\Omega_R} = \left.\frac{\partial \chi^{(1)}(\omega_{pr})}{\partial Q_R}\right|_{Q_R=0}$ is the Raman susceptibility tensor, this time considered for $\omega_{pr}$. The linear susceptibility is modulated by the oscillating phonon only if the mode is Raman-active, i.e. if the second term of Equation (S19) is nonzero. We are therefore only sensitive to Raman-active modes in our detection scheme. These can be purely Raman-active, or both IR- and Raman-active in an inversion symmetry-broken material, such as the 0.9 THz and 1.4 THz modes considered in our work. The nonequilibrium polarization $\boldsymbol{P}_{pr}(Q_R, \boldsymbol{E}^{pr})$ induced in the material in the presence of the probe field, written as a function of the coherent phonon vibrational coordinate is therefore:

$$\boldsymbol{P}_{pr}(Q_R, \boldsymbol{E}^{pr}) = \varepsilon_0 \chi^{(1)}(\omega_{pr}, Q_R) \cdot \boldsymbol{E}^{pr} = \chi_0^{(1)}(\omega_{pr})\boldsymbol{E}^{pr} + \varepsilon_0 Q_R \boldsymbol{R}^{\omega_{pr},\Omega_R} \boldsymbol{E}^{pr} \tag{S20}$$

The first term of Equation (S20) includes the linear, as well as pump field-induced nonlinear polarization, related to the nonresonant electronic response. In the second term of the right-hand side of Equation (S20), the phonon amplitude $Q_R$ is a function of the driving field and corresponds to an effective nonlinear polarization. The solutions for the coherent phonon equations of motion (S14) and (S18) in the case of linear and Raman-type driving give us, respectively:

$$Q_{\text{IR}}(\omega = \omega_1) = \frac{F_{\text{dr}}^{\text{IR}}(\omega)}{\Omega_{\text{IR}}^2 - \omega^2 - i\Gamma\omega} = \frac{\mathbf{Z}^* \cdot \mathbf{E}^{\text{dr},1}(\omega_1)}{\Omega_{\text{IR}}^2 - \omega^2 - i\Gamma\omega}, \tag{S21}$$

$$Q_{\text{R}}(\omega = \omega_1 \pm \omega_2) = \frac{F_{\text{dr}}^{\text{R}}(\omega)}{\Omega_{\text{R}}^2 - \omega^2 - i\Gamma\omega} = \frac{\varepsilon_0 \mathbf{E}^{\text{dr},2}(\omega_2) \cdot \mathbf{R}^{\omega_1,\Omega_{\text{R}}} \mathbf{E}^{\text{dr},1}(\omega_1)}{\Omega_{\text{R}}^2 - \omega^2 - i\Gamma\omega}. \tag{S22}$$

From Equations (S20), (S21) and (S22) we obtain the expressions for the *phonon-modulated component of the nonlinear polarization* for the case of IR-driving and Raman driving, $\mathbf{P}_{\text{pr}}^{(2)}(\omega, \omega_{\text{pr}})$ and $\mathbf{P}_{\text{pr}}^{(3)}(\omega_1, \omega_2, \omega_{\text{pr}})$, respectively:

$$\mathbf{P}_{\text{pr}}^{(2)}(\omega, \omega_{\text{pr}}) \propto (\mathbf{Z}^* \cdot \mathbf{E}^{\text{dr},1}(\omega))(\mathbf{R}^{\omega_{\text{pr}},\Omega_{\text{IR,R}}} \mathbf{E}^{\text{pr}}), \tag{S23}$$

$$\mathbf{P}_{\text{pr}}^{(3)}(\omega_1, \omega_2, \omega_{\text{pr}}) \propto (\varepsilon_0 \mathbf{E}^{\text{dr},2}(\omega_2) \cdot \mathbf{R}^{\omega_1,\Omega_{\text{R}}} \mathbf{E}^{\text{dr},1}(\omega_1))(\mathbf{R}^{\omega_{\text{pr}},\Omega_{\text{IR,R}}} \mathbf{E}^{\text{pr}}). \tag{S24}$$

$\Omega_{\text{IR,R}}$ is the frequency of a mode which is simultaneously IR and Raman-active, enabling both linear driving and Raman-type probing. $\Omega_{\text{R}}$ is the frequency of a Raman-driven mode. The polarization is second-order $\mathbf{P}_{\text{pr}}^{(2)}(\omega, \omega_{\text{pr}})$ in the case of linear driving of IR-and-Raman-active modes and third-order $\mathbf{P}_{\text{pr}}^{(3)}(\omega_1, \omega_2, \omega_{\text{pr}})$ for Raman-type driving. In the wave-mixing picture, we can introduce the effective nonlinear susceptibility tensors $\chi_{\text{eff}}^{(2)}$ and $\chi_{\text{eff}}^{(3)}$:

$$P_{\text{pr},i}^{(2)} = \chi_{\text{eff},ijk}^{(2)} E_j^{\text{pr}} E_k^{\text{dr},1} \propto \mathbf{R}_{ij}^{\omega_{\text{pr}},\Omega_{\text{IR,R}}} E_j^{\text{pr}} Z_k^* E_k^{\text{dr},1}, \tag{S25}$$

$$P_{\text{pr},i}^{(3)} = \chi_{\text{eff},ijkl}^{(3)} E_j^{\text{pr}} E_k^{\text{dr},1} E_l^{\text{dr},2} \propto \mathbf{R}_{ij}^{\omega_{\text{pr}},\Omega_{\text{R}}} \mathbf{R}_{kl}^{\omega_1,\Omega_{\text{R}}} E_j^{\text{pr}} E_k^{\text{dr},1} E_l^{\text{dr},2}, \tag{S26}$$

where the repeating indices indicate summation.

The effective $\chi_{\text{eff}}^{(3)}$ is constructed as a product of two standard Raman susceptibility tensors, provided the absence of absorption (transparency at the probe frequency).[24] The nonlinear polarization is the source of signal fields $\mathbf{E}^{\text{sig}}$ at frequencies shifted from the probe frequency by the phonon frequency $\Omega$: ($\omega_{\text{pr}} \pm \Omega$). Our experiment is equivalent to heterodyne detection of signal fields generated in coherent anti-Stokes Raman scattering (CARS) and coherent Stokes Raman scattering (CSRS) processes,[25,26] with the transmitted probe field acting as the local oscillator in the heterodyne detection scheme.

*S17.3. Signal field detection in balanced detection scheme*

In our experiment, the signal fields co-propagate together with the remaining transmitted field $E^{\mathrm{pr,tr}}$ towards the detection setup. Because $E^{\mathrm{sig}}$ frequencies lie within the spectral bandwidth of the fs probe pulse (∼ 25 THz), we can consider the superposition of the signal and transmitted probe field as an effective change of the probe beam polarization state (transient birefringence picture[27]). The balanced detection scheme consists of a quarter- (QWP) and half-wave plate (HWP). During balancing, the HWP is first adjusted so that the static transmitted probe field amplitudes along the $x$ and $y$ directions are equal and the QWP is afterwards added and then adjusted to compensate for any static ellipticity. In this way, we afterwards convert the THz-induced elliptical polarization of the transmitted probe beam to a rotation of its linear polarization. The sum of the signal field and probe field transmitted through the sample ($E^{\mathrm{sig}} + E^{\mathrm{pr,tr}}$) passes through a Wollaston prism which projects the field components polarized along the $x$ and $y$ directions to two separate photodiodes. We use ~~and~~ an incident probe polarization angle such that in the balanced condition in the absence of THz modulation,ced probe polarization angle such that in the balanced condition in the absence of THz modulation, the transmitted probe field is along the $(-x, y)$ direction after the two waveplates.

The detected transient birefringence signal $S$ is proportional to the intensity difference on the two diodes $S \propto (E^{\mathrm{sig}} + E^{\mathrm{pr,tr}})_x^2 - (E^{\mathrm{sig}} + E^{\mathrm{pr,tr}})_y^2$. In the balanced case $E_x^{\mathrm{pr,tr}} = E_y^{\mathrm{pr,tr}}$ and assuming the quadratic terms $(E_x^{\mathrm{sig}})^2$ and $(E_x^{\mathrm{sig}})^2$ are negligibly small, $S \propto (E_x^{\mathrm{sig}} - E_y^{\mathrm{sig}})$.

As seen from the above expression, the signal $S$ scales linearly with the amplitude of the nonlinear signal field. The phonon-modulated nonequilibrium polarization, which is the source of the signal field, in turn scales linearly with the coherent phonon amplitude (Equation S20). In the time domain the signal $S(t)$ therefore directly visualizes the instantaneous periodic displacement of the lattice mode $Q(t)$.

**Section S18. Numerical simulation of the signal angle dependence**

In our numerical calculation, we use the Jones formalism to implement the probe and pump electric field polarization vectors and optical elements. We use electric field vectors normalized to unity, and define the matrix describing the effect of a QWP at the angle corresponding to our lab geometry as (for simplicity here neglecting adjustments of the optics to compensate for static birefringence):

$$\mathbf{M}_{\mathrm{QWP}}^{45°} = \frac{1}{\sqrt{2}}\begin{bmatrix} 1 & i \\ i & 1 \end{bmatrix}. \tag{S29}$$

Similarly, we implement the HWP matrix:

$$\mathbf{M}_{\mathrm{HWP}}^{45°} = \begin{bmatrix} 0 & -1 \\ -1 & 0 \end{bmatrix}, \tag{S30}$$

and on describing the Wollaston prism:

$$\mathbf{M}_{\mathrm{WP}}^{0°} = \begin{bmatrix} 1 & 0 \\ 0 & -1 \end{bmatrix}. \tag{S31}$$

To simulate the azimuthal angle dependence experiment, we assume incident THz and probe fields propagating along $z$, with polarization identical to our experimental conditions:

$$\boldsymbol{E}^{\mathrm{THz}} = \begin{bmatrix} 0 \\ 1 \end{bmatrix}, \qquad \boldsymbol{E}^{\mathrm{pr}} = \frac{1}{\sqrt{2}} \begin{bmatrix} -1 \\ 1 \end{bmatrix} \tag{S32}$$

We define the Raman tensors for the different vibrational modes:

$$\mathbf{R} = \begin{bmatrix} a & b \\ b & c \end{bmatrix} \tag{S33}$$

under the assumption that ($|a|, |b|, |c| \ll 1$), consistent with the fact that Raman scattering is generally a weak process and only considering the $xy$ plane. We calculate the Raman-type driving force $F^{\mathrm{dr}}(\psi)$ for every sample rotation angle $\psi$ based on the rotated Raman tensor $\mathbf{R}^{\mathrm{dr}}(\psi)$:

$$\mathbf{R}(\psi) = \mathcal{R}(-\psi)\mathbf{R}^{\mathrm{dr}}(\psi = 0)\mathcal{R}(\psi) \tag{S34}$$

where $\mathcal{R}(\psi)$ is the rotation matrix about the $z$ axis by an angle $\psi$ (assuming $\psi = 0$ for the crystal axis along the $y$ direction):

$$\mathcal{R}(\psi) = \begin{bmatrix} \cos(\psi) & \sin(\psi) \\ -\sin(\psi) & \cos(\psi) \end{bmatrix} \tag{S35}$$

The driving force for a given mode is then calculated as:

$$F^{\mathrm{dr}}(\psi) = \boldsymbol{E}^{\mathrm{THz}} \mathbf{R}^{\mathrm{dr}}(\psi) \boldsymbol{E}^{\mathrm{THz}} \tag{S36}$$

We calculate the signal field emitted by the nonlinear polarization $\boldsymbol{E}_{\mathrm{sig}}^{(3)}(\psi)$ as:

$$\boldsymbol{E}_{\mathrm{sig}}^{(3)}(\psi) = e^{-i\pi/2} F^{\mathrm{dr}}(\psi) \tag{S37}$$

The emitted signal field is phase-shifted by $-\pi/2$ relative to $E_{\mathrm{probe}}$. This follows from the one-dimensional inhomogeneous wave equation, considering propagation along $z$, where the field emitted by the nonlinear polarization $P_{\mathrm{NL}}$ is given by:[28]

$$\frac{d\boldsymbol{E}_{\mathrm{sig}}(\omega_{\mathrm{pr}} \pm \Omega; z)}{dz} \propto i P_{\mathrm{NL}}(z), \tag{S38}$$

meaning that the signal field has a $-\pi/2$ phase shift relative to $E_{\mathrm{pr}}$:

$$\boldsymbol{E}_{\text{sig}} \propto e^{-\frac{i\pi}{2}} F^{\text{dr}}(\psi)\, \mathbf{R}^{\text{pr}} \boldsymbol{E}^{\text{pr}} \tag{S39}$$

$\mathbf{R}^{\text{pr}}$ is the Raman tensor describing the probing process. We approximate the total field after the sample as the sum of the incident probe field and the emitted signal field $\boldsymbol{E}^{\text{tot}}(\psi) = (\boldsymbol{E}^{\text{pr}} + \boldsymbol{E}^{(3)}_{\text{sig}}(\psi))$, which after passing through the balancing optics becomes:

$$\boldsymbol{E}^{\text{det}}(\psi) = \mathbf{M}^{0°}_{\text{WP}} \mathbf{M}^{45°}_{\text{HWP}} \mathbf{M}^{45°}_{\text{QWP}} \boldsymbol{E}^{\text{tot}}(\psi) \tag{S40}$$

The calculated intensity of the $x$ and $y$ polarization components on the photodiodes after the Wollaston prism are $I_x = \boldsymbol{E}^{\text{det}}_x \cdot (\boldsymbol{E}^{\text{det}}_x)^*$ and $I_y = \boldsymbol{E}^{\text{det}}_y \cdot (\boldsymbol{E}^{\text{det}}_y)^*$ and we therefore compute the normalized signal finally as:

$$S(\psi) = (I_x - I_y) \tag{S41}$$

We note that in our calculations the field vectors are normalized and the absolute values of the Raman tensor elements are arbitrary (while fulfilling the condition $\ll 1$). Therefore, we always obtain only the relative sign and magnitude of the driving force, signal field and measured transient birefringence signal as a function of the rotation angle $\psi$ for a particular mode. While this information is sufficient for our symmetry analysis, we cannot conclude on the differences in absolute phonon amplitudes or signal magnitudes between the different modes.

**References**

[1]  M. Frenzel, M. Cherasse, J. M. Urban, F. Wang, B. Xiang, L. Nest, L. Huber, L. Perfetti, M. Wolf, T. Kampfrath, X. Y. Zhu, S. F. Maehrlein, *Sci. Adv.* **2023**, *9*, eadg3856.

[2]  S. F. Maehrlein, P. P. Joshi, L. Huber, F. Wang, M. Cherasse, Y. Liu, D. M. Juraschek, E. Mosconi, D. Meggiolaro, F. De Angelis, X. Y. Zhu, *Proc. Natl. Acad. Sci. U. S. A.* **2021**, *118*, e2022268118.

[3]  L. Huber, S. F. Maehrlein, F. Wang, Y. Liu, X. Y. Zhu, *J. Chem. Phys.* **2021**, *154*, 094202.

[4]  S. G. Motti, M. Kober-Czerny, M. Righetto, P. Holzhey, J. Smith, H. Kraus, H. J. Snaith, M. B. Johnston, L. M. Herz, *Adv Funct Mater.* **2023**, *33*, 2300363.

[5]  X. Chen, H. Lu, Z. Li, Y. Zhai, P. F. Ndione, J. J. Berry, K. Zhu, Y. Yang, M. C. Beard, *ACS Energy Lett.* **2018**, *3*, 2273.

[6]  M. Frenzel, J. M. Urban, L. Nest, T. Kampfrath, M. S. Spencer, S. F. Maehrlein, *Optica* **2024**, *11*, 362.

[7]  M. Sajadi, M. Wolf, T. Kampfrath, *Opt. Express* **2015**, *23*, 28985.

[8]  A. A. Lanin, I. V. Fedotov, A. B. Fedotov, D. A. Sidorov-Biryukov, A. M. Zheltikov, *Sci. Rep.* **2013**, *3*, 1842.

[9]  T. Kohmoto, M. Masui, M. Abe, T. Moriyasu, K. Tanaka, *Phys. Rev. B - Condens. Matter Mater. Phys.* **2011**, *83*, 064304.

[10]  S. Grisard, A. V. Trifonov, I. A. Solovev, D. R. Yakovlev, O. Hordiichuk, M. V. Kovalenko, M. Bayer, I. A. Akimov, *Nano Lett.* **2023**, *23*, 7397.

[11]  M. Hase, K. Mizoguchi, H. Harima, S. ichi Nakashima, K. Sakai, *Phys. Rev. B - Condens. Matter Mater. Phys.* **1998**, *58*, 5448.

[12]  M. Balkanski, R. F. Wallis, E. Haro, *Phys. Rev. B* **1983**, *28*, 1928.

[13]  P. G. Klemens, *Phys. Rev. B* **1975**, *11*, 3206.

[14]  A. Cuquejo-Cid, A. García-Fernández, C. Popescu, J. M. Bermúdez-García, M. A. Señarís-Rodríguez, S. Castro-García, D. Vázquez-García, M. Sánchez-Andújar, *iScience* **2022**, *25*, 104450.

[15]  M. Hase, K. Ishioka, M. Kitajima, K. Ushida, S. Hishita, *Appl. Phys. Lett.* **2000**, *76*, 1258.

[16]  R. Cuscó, E. Alarcón-Lladó, J. Ibáñez, L. Artús, J. Jiménez, B. Wang, M. J. Callahan, *Phys. Rev. B - Condens. Matter Mater. Phys.* **2007**, *75*, 165202.

[17]  B. K. Ridley, *J. Phys. Condens. Matter* **1996**, *8*, 8.

[18]  Z. Zhang, J. Zhang, Z.-J. Liu, N. S. Dahodl, W. Paritmongkol, N. Brown, Y.-C. Chien, Z. Dai, K. A. Nelson, W. A. Tisdale, A. M. Rappe, E. Baldini, **2023**, eadg4417.

[19]  P. Ghosez, J. Michenaud, X. Gonze, *Phys. Rev. B - Condens. Matter Mater. Phys.* **1998**, *58*, 6224.

[20]  D. M. Juraschek, S. F. Maehrlein, *Phys. Rev. B* **2018**, *97*, 174302.

[21]  G. Khalsa, N. A. Benedek, J. Moses, *Phys. Rev. X* **2021**, *11*, 21067.

[22]  S. Maehrlein, A. Paarmann, M. Wolf, T. Kampfrath, *Phys. Rev. Lett.* **2017**, *119*, 127402.

[23]  D. M. Juraschek, D. S. Wang, P. Narang, *Phys. Rev. B* **2021**, *103*, 174302.

[24]  T. E. Stevens, J. Kuhl, R. Merlin, *Phys. Rev. B - Condens. Matter Mater. Phys.* **2002**, *65*, 144304.

[25]  H. Rigneault, P. Berto, *APL Photonics* **2018**, *3*, 091101.

[26]  R. A. Bartels, D. Oron, H. Rigneault, *JPhys Photonics* **2021**, *3*, 042004.

[27]  M. Frenzel, M. Cherasse, J. M. Urban, F. Wang, B. Xiang, L. Nest, L. Huber, L. Perfetti, M. Wolf, T. Kampfrath, X. Y. Zhu, S. F. Maehrlein, *Sci. Adv.* **2023**, *9*, eadg3856.

[28]  R. Boyd, *Nonlinear Optics*, 3rd ed., Academic Press, Burlington, MA, USA, **2008**.